\documentclass[12pt,preprint]{aastex}

\shorttitle{Substellar members in the lambda Orionis cluster}
\shortauthors{Barrado y Navascu\'es et al.}

\begin{document}

\title{The substellar population of the young cluster lambda Orionis}


\author{D. Barrado y Navascu\'es\altaffilmark{1},\altaffilmark{2}}
\affil{Laboratorio de Astrof\'{\i}sica Espacial y F\'{\i}sica Fundamental,
INTA, P.O. Box 50727, E-2808 Madrid, SPAIN}
\email{barrado@laeff.esa.es}

\author{J.R. Stauffer\altaffilmark{2}}
\affil{IPAC, California Institute of Technology, Pasadena, CA 91125, USA}

\author{J. Bouvier\altaffilmark{3}}
\affil{Laboratoire d'Astrophysique, Observatoire de Grenoble, 
            Universit\'{e} Joseph Fourier, B.P.  53, 38041
            Grenoble Cedex 9,            France}

\author{R. Jayawardhana}
\affil{Department of Astronomy, University of Michigan, 
            830 Dennison Building, Ann Arbor, MI 48109, USA}

\author{Jean-Charles Cuillandre}
\affil{CFHT, 65-1238 Mamalahoa Highway, Kamuela, HI 96743, Hawaii, USA}

\altaffiltext{1}{Visiting Astronomer, Las Campanas Observatory.}
\altaffiltext{2}{Visiting Astronomer, Keck Observatory.}
\altaffiltext{3}{Visiting Astronomer, CFHT Observatory.}


\begin{abstract}
By collecting optical and infrared photometry and low resolution 
spectroscopy, we have identified a large number of  low mass
stars and brown dwarf candidates
 belonging to the young cluster ($\sim$5 Myr) associated
with the binary star $\lambda$ Orionis.
The lowest mass object found is a M8.5 with an estimated mass 
of 0.02 M$_\odot$
($\sim$0.01 M$_\odot$ for objects without spectroscopic confirmation).
 For those objects with spectroscopy, the
measured strength of the H$\alpha$ emission line follows a 
distribution similar to other clusters with the same age range,
with larger equivalent widths for cooler spectral types. 
Three of the brown dwarfs have H$\alpha$ emission equivalent widths
of order 100 \AA, suggestive that they may have accretion disks and
thus are the substellar equivalent of Classical T Tauri stars.
We have derived the Initial Mass Function for the cluster. 
For the substellar regime, the index of the mass spectrum is
$\alpha$=0.60$\pm$0.06, very similar to other young associations.

\end{abstract}





\section{Introduction}

The Lambda Ori OB-T association, located at 400 pc (Murdin \& Penston 1977) 
is a young stellar group which 
has not been studied so far in great depth. 
It is located  inside a fossil giant 
molecular cloud.   The O8 III star $\lambda$$^1$ Ori,
and to a lesser extent the 11 B stars near to it, excite
the  H{\sc II} region S\,264.  Making use of the Infrared Astronomical 
Satellite (IRAS), Zhang et al. (1989) detected
 a dust ring with a diameter
of 9 deg centered around the star $\lambda$ Ori.
This ring is complementary to a shell of neutral hydrogen 
discovered previously by Wade (1957, 1958).
There are two nearby dark clouds within this ring, 
namely B35 and B30, separated from $\lambda$ Ori by
2.2 and 2.7 deg, respectively. 
Based on a H$\alpha$ emission survey, Duerr, Imhoff \& Lada (1982), 
identified three  stellar clusters
centered around B30, B35 and $\lambda$$^1$ Ori, respectively.
Those clusters were
later confirmed from a statistical point of view by 
G\'omez \& Lada (1998).

Dolan \& Mathieu (1999, 2001, 2002) collected moderately deep photometry 
(VRI filters) in an area about 8 sq.deg. centered on the OB association,
 discovering a significant population of low mass stellar members, and
 obtained  medium resolution multifiber spectroscopy for those 
candidates closest to the central star.
Their derived distance of 450$\pm$50 pc is larger than
both the distance derived by Murdin \& Penston (1977) and
the value derived by
Hipparcos (Perryman et al. 1997) for the five stars in the central area, 
380$\pm$30 pc.  According to Dolan and Mathieu (2002),
the turn-off age for the massive stars is of order 6 Myr
(see also Murdin \& Penston 1977 for another age determination based on the 
masive stars, 4 Myr), although the star formation history might be more complex
(Dolan \& Mathieu 2001).

In this paper, we present additional, much deeper photometry, well 
beyond the hydrogen burning  limit at
 0.072 M$_\odot$ (Baraffe et al. 1998).  For 
some of the new candidate members, we have also obtained low resolution
spectroscopy, which allow us to add additional clues 
about their membership and their substellar nature.
In our study, we will assume: an age  of 5 Myr, a distance
400 pc --$(m-M)_0$=8.010-- and 
a reddening of E$(B-V)$=0.12 (Diplas \& Savage 1994).
Section 2 deals with the optical search and  the near infrared counterparts
in the 2MASS All Sky Survey. Section 3 presents the analysis of these 
datasets, whereas the results are summarized in section 4.


\section{The data}

\subsection{Optical survey}

On September 29th and 30th, 1999, we conducted an 
optical photometric survey around the  star
$\lambda$ Orionis, using the CFHT 12K mosaic camera
and the Cousins $R$ and $I$ filters.
This instrument covers a projected area on the sky 
of 42$\times$28 arcminutes,
with a scale of about 0.2 arcsec/pix. 
The data were collected under photometric conditions.
The reduction was performed with the standard pipeline, and the extraction 
of the aperture photometry was carried out within the 
IRAF\footnote{IRAF is distributed by National Optical Astronomy Observatories,
 which is operated by the Association of Universities for Research in
 Astronomy, Inc., under contract to the National Science Foundation, USA}
 environment. The astrometric calibration was tied to the 
USNO-A2.0 catalog (Monet et al. 1998). Our derived coordinates, on average, 
should be accurate to better than 1 arcsec, except for the objects 
closer to the edges of each detector .

Based on the results of the wide-field imaging survey of
Dolan \& Mathieu (2002) --specifically their Figures 6 and 8,
the separation of Lambda Ori from the B30 and B35 clusters, 
and the FOV of the 12K mosaic, we believe that our survey
should mostly include young stars from the cluster associated
with Lambda Ori.  Furthermore, because of the very young age
and moderate distance to the cluster, we expect the very low
mass cluster stars to be well separated from field stars in
a color-magnitude diagram.   Figure 1 illustrates
the location of $\lambda$ Orionis, the dark clouds and 
the CFHT 12K field.

We collected two different sets of exposures, with short and long
exposure times. In the first case, we exposed during 10 seconds 
for each filter, whereas the exposure times were 600 and 900 seconds
 --Cousins $I$ and $R$ filters, respectively, for the second set.
With this method, we increased the dynamical range, and 
our photometry overlaps with previously published data in these filters.
The zero points of the photometric calibration were derived using the
data from Dolan \& Mathieu (1999, 2001, 2002).
Each CCD was calibrated independently. The errors can be estimated
as 0.06 and 0.08 magnitudes, for the shallow and deep exposures,
 respectively.

Figure 2 displays the optical color-magnitude diagram (CMD).
Dots  represent the photometry of field objects (deep images), 
whereas solid circles correspond to our selected candidate 
members (deep and shallow images).
 We have also added other possible, brighter members from 
Dolan \& Mathieu (1999) as crosses. 
The thick, solid line corresponds to an empirical ZAMS from
 Barrado y Navascu\'es et al. (2001b), 
whereas a 5 Myr isochrone  from Baraffe et al. (1998)
 is displayed as a dotted line. This last isochrone, 
 together with the brigther data by Dolan \& Mathieu (1999), was used
 as   the reference  --taken into account photometric errors-- 
for the selection of  candidate members. 
A older isochrone (e.g., 8 Myr) would not
modify in a significant manner our member selection.
 We used  a reddening of  E$(R-I)_C$=0.084.  Visual inspection of the 
figure indicates that there is a clear separation between 
the field population and the cluster sequence, as we had expected.
Hence, we expect contamination of our cluster membership list from field stars
to be relatively small.
Table 1 lists our candidate members, extracted from both the shallow and 
the deep datasets. Columns 4--7 list the optical photometry derived from
the shallow exposures, whereas columns 8--11 contains the data from the 
deep  images.

Our search has a magnitude limit of $I$(lim)$\sim$24.0 mag.
Completeness is achieved at $R$(complete)$\sim$22.75 mag
and $I$(complete)$\sim$22.75, based on the drop (Figure 3)
 in the number of detected
objects for each interval in the $Ic$ and $Rc$ filters 
(Wainscoat et al. 1992; Santiago et al. 1996).
 For cluster members, the faint limit 
is set by $R$(complete) $\sim$ 22.75 at $(R-I)$ = 2.5, corresponding
then to $I$(complete,cluster) $\sim$ 20.2 mag.

\subsection{2MASS infrared photometry}

The limiting magnitude of the 2MASS All Sky release
(Cutri et al$.$ 2003) is about $Ks$$\sim$15.5 mag.
Using a Baraffe et al. (1998) 5 Myr isochrone, 
and the distance and reddening of the Lambda Orionis
cluster, this value corresponds to cluster members
having a mass similar to 0.025 M$_\odot$ ($Ic$$\sim$19.1 mag).
Therefore, we expect that most of our optical candidate 
members, except the faintest, 
 should have a counterpart in the 2MASS catalog.
Our faintest candidates, with $Ic$$\sim$22, would have
masses of order 0.010 M$_\odot$ if they are cluster members
and have an age equal to 5 Myr, according to the Baraffe et al. (2003)  isochrones,
 the so called COND models.
We have matched both datasets using a 5 arcsec radius, much larger
than the combined errors of the  2MASS coordinates  and our own positions. 
Since the surface density is not very large in this field, 
the object identification is generally unambiguous.
2MASS photometry, together with the coordinates
(more accurate than our initial values), are provided in Table 2.
This table lists the $Ic$ magnitude, the distance between the 
optical and the IR source and the 2MASS coordinate, and
 the IR photometry and uncertainties.
Column 12 describes whether the candidate 
fulfills different membership criteria, and the last column
contains our final membership assessment (see section 3.3).

\subsection{Low resolution spectroscopy}

	We have collected low resolution optical spectroscopy in
 two different campaigns. The first one took place in November 3-5, 2002,
at the Keck I telescope.  We made use of LRIS with the 400 l/mm grating 
with a one arcsec slit. 
Typical integration times were 300-900 seconds.  The data 
were processed within the IRAF environment in a standard way.
The wavelength calibration is better than 0.4 \AA,
spectral coverage is 6250--9600 \AA{ }  and the resolution is R$\sim$1100
(6.0 \AA{ } around H$\alpha$),   as measured in a NeAr lamp.
For the second run, we used the B\&C spectrograph attached to the Magellan II
telescope in March 9-11, 2003.
 The reduction was carried out in a similar way as in the case of
the Keck sample. The Magellan spectra have slightly worse
 resolution and larger  spectral range (R$\sim$800, 5000-10200 \AA),
 since in this case we used the 300 l/mm grating.
Relative flux calibration for the spectra was derived using observations
obtained of several spectrophotometric standards.



In total, we obtained spectra for
33 objects out of the 170 possible members  discovered in 
our optical survey. The spectra are displayed 
in Figure 4a-d. Panels a, b and c correspond to 
data from Keck~I, whereas panel d contains the
 Magellan spectra.

\section{Analysis}

\subsection{Spectral types}

Simultaneously to our observations of Lambda Orionis candidate members, 
we observed a large number of  cool stars of different luminosity classes
(V, IV and III) and spectral types ranging from K7 to M9.
These spectra were reduced in the same way and were used to measure
spectral indices as defined by Mart\'{\i}n et al. (1996, 1999)
 and Kirkpatrick  et al. (1999). 
Then, we calibrated these indices with the known spectral types
of this dataset, and derived spectral types for our candidate members.
A final visual inspection was carried out, by  comparing the 
spectrum of the cluster candidate member with the one corresponding 
to a field star of the same spectral type. The final values are
listed in Table 3. 
The error in the spectral  type assignation, except in one case
(LOri-CFHT-165), can be estimated  as half a subclass.

\subsection{Color - color, color-magnitude and 
magnitude  versus spectral type diagrams}

Figures 5a and 5b display optical-infrared color-magnitudes 
diagrams, whereas color-color diagrams are illustrated in 
Figures 6a, 6b and 6c. 
In these diagrams, we have overplotted 5 Myr isochrones
by Baraffe et al. (1998, 2002) and Chabrier et al. (2000),
shifting them to a distance of 400 pc, and using 
the interstellar extinctions of
$A_R$=0.307, $A_I$=0.223, 
$A_J$=0.106, $A_H$=0.066, and $A_K$=0.04.
Additionally, the $Ic$ and $Ks$ magnitudes versus
the spectral type are shown in Figures 7a and 7b.
In these last two  cases, we made used of temperature 
scales by Luhman (1999) -dotted lines-- and  Basri et al. (2000) 
-solid line--for the 
conversion between effective temperatures and spectral types.
Note that Luhman' scales were derived for giant and  dwarfs,
and for an intermediate gravity, which roughly describes
the location of Lambda Ori cluster members, whose gravity should be 
log g$\simeq$4.0, according to Baraffe et al. (1998) models.
In all these figures, different symbols represent
our membership assignments (see next subsection).

\subsection{Membership}

We have established a membership status based on two main criteria:
the spectroscopic and photometric information. When the first one 
was available, we relied mainly on it, otherwise we made use
 of the optical-infrared photometry.

1.- Candidates with spectroscopy.

1.a.- Both spectral type and photometric
data in agreement with cluster sequence. Probable members. They appear
with the label ``Mem+'' in Table 2, and displayed in Figures 5-9
as solid circles.

1.b.- Spectral type compatible with cluster 
sequence, but the location in one or two CCD or CMD does not correspond
to a member. Possible members. Label ``Mem?'' (solid triangles).

1.c.- The estimated spectral type does not agree with the photometry, but
 all CMD and CCD indicate membership.
We also have included here LOri-CFHT-119, with photometry in disagreement with 
membership and whose spectral type does not indicate clearly its status.
Possible non-member. ``NM?'' (open triangles).
 
1.d. -Both spectral type and photometric
data in disagreement with cluster sequence. Probable non-members.
``NM+'' (open circles).

2.- Candidates without spectroscopy.

2.a.- All CMD and CCD indicate membership. Probable members.
They are listed in Table 2 with the tag ``Mem''. Shown as 
plus symbols  in Figures 5-9.

2.b.- One CMD or CCD in disagreement with the cluster 
sequence. Possible members. ``Mem?'' (crosses).

2.c.- At least a CMD and another CMD or CCD in disagreement 
with the membership. Possible non-member. ``NM-'' (small dots).

2d.- No information in the 2MASS All Sky release. Most of them 
are too faint to have been detected by that survey. They are
labeled with the tag ``???''. 

One object, LOri-081, is located in Figure 5b in a position which might 
indicate  a large reddening. However, Figures 6a and 6b suggest that this 
object has a  near infrared excess which is characteristic of Classical 
TTauri stars. Note, however, that it H$\alpha$ equivalent width is not
 very large (see section 3.5).
 Since this low mass star has no other indicator which might
suggest it is a non-member, we have catalogued it as a member.

Figure 8a shows the absolute $Ic$ magnitude against the dereddened  
$(R-I)c$ color and a comparison with theoretical isochrones with ages in the
range 1-10 Myr by Baraffe et al. (1998) --NextGen models. 
Figure 8b displays the same set of theoretical models (isochrones and 
evolutionary tracks) in a HR diagram. 
Luminosities were derived  from either  the $Ic$ or the $Ks$
magnitudes --open squares or stars, respectively-- and bolometric corrections
by Comer\'on et al. (2000) and Tinney et al. (1993).
Effective temperatures come from the Luhman's scale (1999) for
intermediate gravity. This figure indicates that 
the age of the cluster is bracketed by 3 and 10 Myr.

Of the 170 candidate members
selected from the (R, R-I) CMD, 24 have no IR data, and of the remaining
146, 104 are classified probable members, 22 as possible members and 20 as
non members. Hence, the contamination level of the optical sample is of
order of 25\% for both subsamples with and without spectral information.
Additionally, a significant fraction of the confirmed members should be, 
based on the assumed distance, reddening  and age ranges,
 bona fide brown dwarfs.

\subsection{Spatial distribution}

We have checked whether there is any concentration of bona fide,
probable or possible  members close to the $\lambda$ Orionis multiple star
(or, conversely, a clustering of non-members). Figure 9 shows the
spatial distribution. We see no conclusive
evidence for clustering or sub-clustering.   Due to the geometry of our survey
(a rectangle of 42$\times$28 armin) and the total area covered 
 by it (smaller than the expected projected total size of the cluster
about 1-2 sq.deg), we have not  
derive the  radial distribution of objects.

\subsection{The Mass Function}

We have derived the cluster's Initial Mass Function (IMF) 
 in the mass range 0.02-1.2
M$_\odot$ (the CFHT 1999 RI survey) and from Dolan \& Mathieu's (1999,
2001) brighter sample over the mass range 0.3-4.7 M$_\odot$. 
We used two sets of data:
In the first one we removed  the probable non-members (NM$+$).
 In the second set, we only retained the probable members 
(``Mem$+$'' and ``Mem'').
Masses were
computed from dereddened I-band  magnitudes, 
using 3,  5  and 10 Myr isochrones
of Baraffe et al.'s (1998) model. Note that, as shown in
Barrado y Navascu\'es et al. (2001b), the use of other models
do not affect the derived IMF in an significant  degree.

The cluster's MF is shown in Figure 10a, where Dolan \& Mathieu's sample
has been restricted to the area in common with the CFHT 1999 survey. The
vertical segment denotes the location of the completeness limit of the CFHT
survey. A power law fit
--carried out with  the sample without the probable non-members--
 to the mass spectrum ($dN/dm \propto m^{-\alpha}$)
indicates an index $\alpha$=+0.60$\pm$0.06 across the stellar/substellar limit
(0.03-0.14 M$_\odot$), and a slightly steeper index $\alpha$=+0.86$\pm$0.05
 over the whole mass range from $\sim$0.024 M$_\odot$ to 0.86 M$_\odot$,
 using a 5 Myr isochrone.
A 5 Myr isochrone from Burrows et al. (1997) gives  $\alpha$=+0.69$\pm$0.17
in the range 0.20--0.015 M$_\odot$, whereas models from D'Antona \& Mazzitelli
(1997, 1998) are almost identical --regarding the power law index--
to those obtained with  Baraffe et al. (1998).  

 On the other hand, 3 and 10 Myr isochrones from Baraffe et al. (1998)
 yield
 $\alpha$=+0.92$\pm$0.04 and  $\alpha$=+0.71$\pm$0.06, respectively
(again, in the range 0.024 M$_\odot$ to 0.86 M$_\odot$).

In the case of second dataset, which only contains bona-fide members,
the slope of the IMF, for a 5 Myr isochrone, is  $\alpha$=+0.57$\pm$0.06.
Note, however, that this is a minimum value, since some among  
the faintest candidate  members do not have IR data (they do not appear
in the 2MASS catalog) and because our spectroscopic survey was biased
(we observed preferably in the range I$c$=15-17).

 The slope of Lambda  Ori MF at lower
masses and into the substellar domain is quite similar to that derived for
other young clusters by some of us,
 e.g. Sigma Orionis ($\alpha$=+0.8, B\'ejar et
al. 2001), Alpha Per ($\alpha$=+0.6, Barrado y Navascu\'es et al. 2002b)
and the Pleiades ($\alpha$=+0.6, Bouvier et al. 1998; Moraux et al. 2003). 
The age of these
clusters is estimated as 5, 80 and 125 Myr, respectively (Zapatero Osorio
et al. 2002; Stauffer et al. 1998, 1999; Barrado y Navascu\'es et
al. 2004). The $\alpha$ index is also similar to the results 
obtained in other stellar associations 
such as Trapezium, IC348 or Taurus (Luhman et al. 2000, 2003;
 Lucas \& Roche 2000; Hillenbrand \& Carpenter 2000;
 Najita et al. 2000; Preibisch et al. 2002; Brice\~no et al. 2002;
 Muench  et al. 2003).

Figure 10b displays the same data but plotted as dN/dlogM (instead of
dN/dM) as a function of mass. The cluster's mass function appears to rise
from 3 to 0.8 M$_\odot$, remains about flat down to 0.1 M$_\odot$, and
decreases toward lower masses and into the substellar regime. This
behaviour is not unlike the lognormal shape of the mass function derived
for both the Pleiades cluster (Moraux et al. 2003), as illustrated on Fig. 10b
with a dashed line,  and the galactic field (Chabrier 2003).

\subsection{H$\alpha$ emission}

We have measured the H$\alpha$ equivalent with 
using the pseudo-continuum. Note that low resolution
spectra tend to produce larger equivalent widths compared with 
data taken at higher resolution. These values, plotted versus 
the derived spectral types with
symbols as in the previous figures, 
 are displayed in Figure 11, where
we also include a comparison with the ``twin'' cluster 
sigma Orionis (Wolk 1996; Walter et al.  1997),
 with approximately the same age and 
 located at similar distance. Sigma Orionis H$\alpha$ equivalent 
widths and spectral types come from 
B\'ejar et al. (1999),
Barrado y Navascu\'es et al. (2001a, 2002a, 2003), 
Zapatero Osorio et al. (2002),
Muzerolle et al. (2003), and
Jayawardhana, Mohanty \& Basri (2003). 
The  dashed lines 
indicate the 5 and 20 \AA{ } criteria which separate Classical -accreting-
 from Weak-line -non accreting- T~Tauri stars, whereas 
the thick line corresponds to the
criterion differentiating accreting from non accreting objects, based on low 
resolution spectroscopy (Barrado y Navascu\'es \& Mart\'{\i}n 2003).
This criterion is based on the upper envelope of chromospheric
H$\alpha$\ emission
present in cluster members
belonging IC2391, Alpha Per and the Pleiades.
Although the low resolution and relatively low S/N of our
spectra --especially at the faint end-- 
do not allow us to definitively prove that any of our
objects have accretion disks (Muzerolle et al. 2003; 
Jayawardhana, Mohanty \& Basri 2003), the
strength of H$\alpha$ for several of the confirmed members of the
Lambda Ori cluster suggests to us that some of them might be accreting.

The three Lambda Ori members with strongest H$\alpha$ emission
are:  LOri 140 (eq. width = 72.8 \AA), LOri 156 (eq. width = 101.7 \AA) and
LOri 161 (eq. width = 123 \AA).   LOri 140 might present [OI]6300
 in emission.   Note that all of the objects with the
largest H$\alpha$ equivalent widths are located in the substellar
regime of the CMD.
However, we do not detect  [OII]7329 \AA, [SII]6717\&6731 \AA{ }
or the CaII IRT, which can be seen in LS-RCrA~1, another brown dwarf
of similar mass and spectral type  (M6.5 IV),
discovered by Fern\'andez \& Comer\'on (2001) and analyzed in depth 
by Barrado y Navascu\'es, Mohanty \& Jayawardhana (2004).
The spectral properties of these three members  are closer
 to those present in
2M1207-3932, a M8 brown dwarf which belong to the
 TW Hydrae association (Mohanty et al. 2003).


\section{Summary and Conclusions.}


We have collected deep optical photometry in about 0.3 sq.deg 
around the binary star $\lambda$ Orionis, extending to well below the
substellar boundary. The combination of this dataset 
set with near infrared photometry and 
low resolution spectroscopy --i.e., spectral types-- 
allow us to cull from the initial membership list 
the possible and probable low mass members of the cluster, both of 
stellar and substellar nature. 
We note, however, that additional work is required, in order to study
other youth indicators such as low-gravity features and  the detection of lithium.
We conclude that the pollution fraction due to
interlopers is low (similar to 25 \% for both the sample with or without 
spectroscopic information).
The faintest object whose membership has been established
is a brown dwarf with a mass slightly below 0.020 M$_\odot$ (based
on the Chabrier and Baraffe models) and a M8.5 
spectral type.
Moreover, H$\alpha$ equivalent widths have been measured 
in the spectra.   A plot of the H$\alpha$\ equivalent widths
as a function of spectral type shows a very similar distribution
for Lambda Ori and for the similar age Sigma Ori clusters,
with an increase
on average for cooler spectral types. Some of the Lambda Orionis stars
and brown dwarfs have W(H$\alpha$) larger
than the chromospheric saturation limit. By  analogy with Classical TTauri
stars, they might have an accretion disk.
We have also derived the Initial Mass Function in the range
4.7-0.02 M$_\odot$, which shows different types of behavior when 
displayed as a mass spectrum.
Across the stellar/substellar boundary, the index of a
power law fit is $\alpha$=+0.60$\pm$0.06,
 quite similar to values recently derived
for other young clusters in the same mass range.


\acknowledgements
We  do appreciate the referee's comments and suggestions (Victor B\'ejar).
 DByN  is  supported by  the Spanish
``Programa Ram\'on y Cajal'', PNAyA2001-1124-C02 and 
PNAyA AYA2003-05355.  This publication makes 
use of data products from the Two Micron All Sky Survey.


\setcounter{table}{0}
\begin{table*}
\tiny
\caption[ ]{Positions and optical photometric data for our lambda Orionis candidate members.}
\begin{tabular}{lllcccccccccc}
\hline
LOri&   R.A      & DEC         &\, & Ic    &(R-I)c&$\delta$R&$\delta$I&\,& Ic   &(R-I)c&$\delta$R&$\delta$I \\ 
\cline{2-3} \cline{5-8} \cline{10-13}
CFHT&  \multicolumn{2}{c}{(2000.0)}  & & \multicolumn{4}{c}{shallow exp.} & & \multicolumn{4}{c}{deep exp.}\\
\#  & (h:m:s)    & ($^\circ$ ' '') &       &      &         &         &      &      &         &          \\&                       \\
\hline
001 & 5:33:47.18 &  9:55:38.5 & &   12.52 &   0.69 &   0.01 &   0.01 & &  ---   &  ---   &  ---   &  ---    \\ 
002 & 5:36:10.36 & 10:08:54.9 & &   12.64 &   0.80 &   0.01 &   0.01 & &  ---   &  ---   &  ---   &  ---    \\ 
003 & 5:35:55.44 &  9:56:31.5 & &   12.65 &   0.74 &   0.01 &   0.01 & &  ---   &  ---   &  ---   &  ---    \\ 
004 & 5:35:47.55 &  9:45:50.5 & &   12.65 &   1.06 &   0.01 &   0.01 & &  ---   &  ---   &  ---   &  ---    \\ 
005 & 5:33:53.65 &  9:43:07.97& &   12.67 &   0.71 &   0.01 &   0.01 & &  ---   &  ---   &  ---   &  ---    \\ 
006 & 5:35:16.26 &  9:55:17.8 & &   12.75 &   0.80 &   0.01 &   0.01 & &  ---   &  ---   &  ---   &  ---    \\ 
007 & 5:34:29.55 &  9:48:58.7 & &   12.78 &   0.94 &   0.01 &   0.01 & &  ---   &  ---   &  ---   &  ---    \\ 
008 & 5:35:57.97 &  9:54:32.8 & &   12.79 &   0.81 &   0.01 &   0.01 & &  ---   &  ---   &  ---   &  ---    \\ 
009 & 5:34:46.34 & 10:06:35.6 & &   12.95 &   0.75 &   0.01 &   0.01 & &  ---   &  ---   &  ---   &  ---    \\ 
010 & 5:34:32.96 & 10:08:41.1 & &   12.96 &   0.74 &   0.01 &   0.01 & &  ---   &  ---   &  ---   &  ---    \\ 
011 & 5:34:44.66 &  9:53:57.5 & &   13.01 &   0.83 &   0.01 &   0.01 & &  ---   &  ---   &  ---   &  ---    \\ 
012 & 5:35:05.95 &  9:52:07.8 & &   13.03 &   0.77 &   0.01 &   0.01 & &  ---   &  ---   &  ---   &  ---    \\ 
013 & 5:33:56.35 &  9:53:56.69& &   13.03 &   1.18 &   0.01 &   0.01 & &  ---   &  ---   &  ---   &  ---    \\ 
014 & 5:36:19.03 & 10:03:50.8 & &   13.03 &   0.81 &   0.01 &   0.01 & &  ---   &  ---   &  ---   &  ---    \\ 
015 & 5:34:21.84 & 10:04:14.5 & &   13.05 &   0.78 &   0.01 &   0.01 & &  ---   &  ---   &  ---   &  ---    \\ 
016 & 5:35:13.50 &  9:55:24.5 & &   13.18 &   0.89 &   0.01 &   0.01 & &  ---   &  ---   &  ---   &  ---    \\ 
017 & 5:36:20.49 &  9:52:19.3 & &   13.19 &   0.80 &   0.01 &   0.01 & &  ---   &  ---   &  ---   &  ---    \\ 
018 & 5:36:16.59 &  9:50:48.8 & &   13.26 &   0.95 &   0.01 &   0.01 & &  ---   &  ---   &  ---   &  ---    \\ 
019 & 5:35:13.69 &  9:56:28.8 & &   13.31 &   1.02 &   0.01 &   0.01 & &  ---   &  ---   &  ---   &  ---    \\ 
020 & 5:34:57.57 &  9:46:07.5 & &   13.31 &   1.34 &   0.01 &   0.01 & &  ---   &  ---   &  ---   &  ---    \\ 
021 & 5:35:06.94 &  9:48:57.8 & &   13.38 &   0.88 &   0.01 &   0.01 & &  ---   &  ---   &  ---   &  ---    \\ 
022 & 5:35:51.35 &  9:55:10.8 & &   13.38 &   1.03 &   0.01 &   0.01 & &  ---   &  ---   &  ---   &  ---    \\ 
023 & 5:35:57.65 &  9:47:34.6 & &   13.44 &   0.99 &   0.01 &   0.01 & &  ---   &  ---   &  ---   &  ---    \\ 
024 & 5:34:57.11 &  9:54:36.1 & &   13.45 &   0.98 &   0.01 &   0.01 & &  ---   &  ---   &  ---   &  ---    \\ 
025 & 5:36:20.18 &  9:44:02.0 & &   13.45 &   0.91 &   0.01 &   0.01 & &  ---   &  ---   &  ---   &  ---    \\ 
026 & 5:34:36.20 &  9:53:43.8 & &   13.47 &   1.10 &   0.01 &   0.01 & &  ---   &  ---   &  ---   &  ---    \\ 
027 & 5:35:11.01 & 10:07:36.7 & &   13.50 &   0.99 &   0.01 &   0.01 & &  ---   &  ---   &  ---   &  ---    \\ 
028 & 5:36:18.85 &  9:51:35.3 & &   13.65 &   1.21 &   0.01 &   0.01 & &  ---   &  ---   &  ---   &  ---    \\ 
029 & 5:35:25.36 & 10:08:38.7 & &   13.69 &   1.20 &   0.01 &   0.01 & &  ---   &  ---   &  ---   &  ---    \\ 
030 & 5:35:12.57 &  9:55:18.8 & &   13.74 &   1.21 &   0.01 &   0.01 & &  ---   &  ---   &  ---   &  ---    \\ 
031 & 5:34:49.02 &  9:58:03.4 & &   13.75 &   1.15 &   0.01 &   0.01 & &  ---   &  ---   &  ---   &  ---    \\ 
032 & 5:36:09.31 &  9:47:03.7 & &   13.80 &   1.24 &   0.01 &   0.01 & &  ---   &  ---   &  ---   &  ---    \\ 
033 & 5:35:34.83 & 10:00:35.1 & &   13.81 &   1.01 &   0.01 &   0.01 & &  ---   &  ---   &  ---   &  ---    \\ 
034 & 5:35:19.92 & 10:02:36.6 & &   13.97 &   1.13 &   0.01 &   0.01 & &  ---   &  ---   &  ---   &  ---    \\ 
035 & 5:35:15.16 & 10:01:07.1 & &   13.97 &   1.28 &   0.01 &   0.01 & &  ---   &  ---   &  ---   &  ---    \\ 
036 & 5:34:39.25 & 10:01:29.4 & &   13.98 &   1.49 &   0.01 &   0.01 & &  ---   &  ---   &  ---   &  ---    \\ 
037 & 5:34:35.57 &  9:59:44.3 & &   13.99 &   1.18 &   0.01 &   0.01 & &  ---   &  ---   &  ---   &  ---    \\ 
038 & 5:33:49.96 &  9:50:37.3 & &   14.01 &   1.09 &   0.01 &   0.01 & &  ---   &  ---   &  ---   &  ---    \\ 
039 & 5:35:57.06 &  9:46:53.0 & &   14.02 &   1.23 &   0.01 &   0.01 & &  ---   &  ---   &  ---   &  ---    \\ 
040 & 5:35:39.49 &  9:50:33.0 & &   14.06 &   1.32 &   0.01 &   0.01 & &  ---   &  ---   &  ---   &  ---    \\ 
041 & 5:35:30.47 &  9:50:34.5 & &   14.10 &   1.45 &   0.01 &   0.01 & &  ---   &  ---   &  ---   &  ---    \\ 
042 & 5:36:07.12 & 10:09:48.0 & &   14.14 &   1.17 &   0.01 &   0.01 & &  ---   &  ---   &  ---   &  ---    \\ 
043 & 5:35:02.73 &  9:56:49.4 & &   14.16 &   1.30 &   0.01 &   0.01 & &  ---   &  ---   &  ---   &  ---    \\ 
044 & 5:34:08.38 &  9:51:25.27& &   14.17 &   1.22 &   0.01 &   0.01 & &  ---   &  ---   &  ---   &  ---    \\ 
045 & 5:35:07.40 &  9:58:23.8 & &   14.23 &   1.33 &   0.01 &   0.01 & &  ---   &  ---   &  ---   &  ---    \\ 
046 & 5:34:26.08 &  9:51:49.7 & &   14.36 &   1.28 &   0.01 &   0.01 & &  ---   &  ---   &  ---   &  ---    \\ 
047 & 5:35:55.64 &  9:50:53.7 & &   14.38 &   1.53 &   0.01 &   0.01 & &  ---   &  ---   &  ---   &  ---    \\ 
048 & 5:35:12.58 &  9:53:10.8 & &   14.41 &   1.37 &   0.01 &   0.01 & &  ---   &  ---   &  ---   &  ---    \\ 
049 & 5:35:01.00 &  9:49:36.4 & &   14.50 &   1.27 &   0.01 &   0.01 & &  ---   &  ---   &  ---   &  ---    \\ 
050 & 5:34:56.39 &  9:55:03.8 & &   14.54 &   1.36 &   0.01 &   0.01 & &  ---   &  ---   &  ---   &  ---    \\ 
\hline
\end{tabular}
\end{table*}

\setcounter{table}{0}
\begin{table*}
\tiny
\caption[ ]{Positions and optical photometric data for our lambda Orionis candidate members.}
\begin{tabular}{lllcccccccccc}
\hline
LOri&   R.A      & DEC         &\, & Ic    &(R-I)c&$\delta$R&$\delta$I&\,& Ic   &(R-I)c&$\delta$R&$\delta$I \\ 
\cline{2-3} \cline{5-8} \cline{10-13}
CFHT&  \multicolumn{2}{c}{(2000.0)}  & & \multicolumn{4}{c}{shallow exp.} & & \multicolumn{4}{c}{deep exp.}\\
\#  & (h:m:s)    & ($^\circ$ ' '') &       &      &         &         &      &      &         &          \\&                       \\
\hline
051 & 5:36:12.14 & 10:00:57.4 & &   14.60 &   1.31 &   0.01 &   0.01 & &  ---   &  ---   &  ---   &  ---    \\ 
052 & 5:34:11.77 &  9:57:03.91& &   14.63 &   1.30 &   0.01 &   0.01 & &  ---   &  ---   &  ---   &  ---    \\ 
053 & 5:34:36.71 &  9:52:58.1 & &   14.72 &   1.36 &   0.01 &   0.01 & &  ---   &  ---   &  ---   &  ---    \\ 
054 & 5:35:52.52 &  9:48:31.6 & &   14.73 &   1.46 &   0.01 &   0.01 & &  ---   &  ---   &  ---   &  ---    \\ 
055 & 5:35:21.43 &  9:49:56.9 & &   14.76 &   1.36 &   0.01 &   0.01 & &  ---   &  ---   &  ---   &  ---    \\ 
056 & 5:34:58.36 &  9:53:46.7 & &   14.87 &   1.56 &   0.01 &   0.01 & &  ---   &  ---   &  ---   &  ---    \\ 
057 & 5:35:11.35 & 10:00:50.8 & &   15.04 &   1.59 &   0.01 &   0.01 & &  ---   &  ---   &  ---   &  ---    \\ 
058 & 5:36:10.37 & 10:00:19.1 & &   15.06 &   1.51 &   0.01 &   0.01 & &  ---   &  ---   &  ---   &  ---    \\ 
059 & 5:34:23.57 &  9:43:43.3 & &   15.10 &   1.47 &   0.01 &   0.01 & &  ---   &  ---   &  ---   &  ---    \\ 
060 & 5:35:20.00 &  9:49:06.6 & &   15.14 &   1.42 &   0.01 &   0.01 & &  ---   &  ---   &  ---   &  ---    \\ 
061 & 5:35:18.19 &  9:52:24.2 & &   15.15 &   1.43 &   0.01 &   0.01 & &  ---   &  ---   &  ---   &  ---    \\ 
062 & 5:35:15.33 &  9:48:37.3 & &   15.16 &   1.46 &   0.01 &   0.01 & &  ---   &  ---   &  ---   &  ---    \\ 
063 & 5:35:19.16 &  9:54:41.9 & &   15.34 &   1.46 &   0.01 &   0.01 & &  ---   &  ---   &  ---   &  ---    \\ 
064 & 5:35:51.99 &  9:50:29.8 & &   15.34 &   1.44 &   0.01 &   0.01 & &  ---   &  ---   &  ---   &  ---    \\ 
065 & 5:35:17.95 &  9:56:58.4 & &   15.37 &   1.52 &   0.01 &   0.01 & &  ---   &  ---   &  ---   &  ---    \\ 
066 & 5:35:06.76 &  9:57:29.5 & &   15.40 &   1.72 &   0.01 &   0.01 & &  ---   &  ---   &  ---   &  ---    \\ 
067 & 5:36:26.37 &  9:45:46.8 & &   15.53 &   1.52 &   0.01 &   0.01 & &  ---   &  ---   &  ---   &  ---    \\ 
068 & 5:34:48.01 &  9:43:25.82& &   15.20 &   1.56 &   0.01 &   0.01 & &  ---   &  ---   &  ---   &  ---    \\ 
069 & 5:34:43.96 &  9:48:35.71& &   15.20 &   1.69 &   0.01 &   0.01 & &  ---   &  ---   &  ---   &  ---    \\ 
070 & 5:36:00.03 & 10:05:49.0 & &   15.61 &   1.57 &   0.01 &   0.01 & &  ---   &  ---   &  ---   &  ---    \\ 
071 & 5:34:15.78 & 10:06:54.44& &   15.45 &   1.64 &   0.01 &   0.01 & &  15.63 &   1.50 &   0.01 &   0.01  \\ 
072 & 5:34:11.34 &  9:42:06.35& &   15.35 &   1.65 &   0.01 &   0.01 & &  ---   &  ---   &  ---   &  ---    \\ 
073 & 5:34:46.82 &  9:50:37.88& &   15.28 &   1.56 &   0.01 &   0.01 & &  ---   &  ---   &  ---   &  ---    \\ 
074 & 5:36:00.57 &  9:42:38.15& &   15.39 &   1.64 &   0.01 &   0.01 & &  ---   &  ---   &  ---   &  ---    \\ 
075 & 5:34:55.22 & 10:00:35.31& &   15.23 &   1.72 &   0.01 &   0.01 & &  ---   &  ---   &  ---   &  ---    \\ 
076 & 5:35:11.00 &  9:57:45.0 & &   15.81 &   1.58 &   0.01 &   0.01 & &  ---   &  ---   &  ---   &  ---    \\ 
077 & 5:34:41.68 &  9:42:41.13& &   15.82 &   1.61 &   0.01 &   0.01 & &  15.89 &   1.56 &   0.01 &   0.01  \\ 
078 & 5:36:16.43 &  9:50:16.42& &   15.79 &   1.54 &   0.01 &   0.01 & &  15.92 &   1.43 &   0.01 &   0.01  \\ 
079 & 5:34:48.27 &  9:59:54.63& &   15.89 &   1.60 &   0.01 &   0.01 & &  16.00 &   1.51 &   0.01 &   0.01  \\ 
080 & 5:35:30.04 &  9:59:25.84& &   15.65 &   1.75 &   0.01 &   0.01 & &  16.01 &   1.50 &   0.01 &   0.01  \\ 
081 & 5:33:56.64 & 10:06:14.70& &   15.98 &   1.63 &   0.01 &   0.01 & &  16.02 &   1.59 &   0.01 &   0.01  \\ 
082 & 5:36:00.80 &  9:52:57.40& &   15.92 &   1.63 &   0.01 &   0.01 & &  16.02 &   1.55 &   0.01 &   0.01  \\ 
083 & 5:35:43.44 &  9:54:26.28& &   15.94 &   1.60 &   0.01 &   0.01 & &  16.02 &   1.54 &   0.01 &   0.01  \\ 
084 & 5:35:50.83 & 10:09:46.32& &   15.78 &   1.62 &   0.01 &   0.01 & &  16.03 &   1.45 &   0.01 &   0.01  \\ 
085 & 5:35:21.54 &  9:53:29.0 & &   16.04 &   1.61 &   0.01 &   0.01 & &  ---   &  ---   &  ---   &  ---    \\ 
086 & 5:34:11.56 &  9:49:15.61& &   ---   &  ---   &  ---   &  ---   & &  16.09 &   1.50 &   0.01 &   0.01  \\ 
087 & 5:34:33.72 &  9:55:33.44& &   16.16 &   1.59 &   0.01 &   0.01 & &  16.09 &   1.45 &   0.01 &   0.01  \\ 
088 & 5:34:49.51 &  9:58:47.86& &   16.04 &   1.72 &   0.01 &   0.01 & &  16.10 &   1.68 &   0.01 &   0.01  \\ 
089 & 5:35:04.00 & 10:07:26.66& &   16.13 &   1.66 &   0.01 &   0.01 & &  16.15 &   1.64 &   0.01 &   0.01  \\ 
090 & 5:36:27.06 &  9:51:35.09& &   16.15 &   1.62 &   0.01 &   0.01 & &  16.17 &   1.60 &   0.01 &   0.01  \\ 
091 & 5:34:35.82 &  9:54:25.99& &   16.16 &   1.83 &   0.01 &   0.01 & &  16.18 &   1.83 &   0.01 &   0.01  \\ 
092 & 5:35:50.95 &  9:51:03.86& &   16.18 &   1.64 &   0.01 &   0.01 & &  16.19 &   1.65 &   0.01 &   0.01  \\ 
093 & 5:34:41.19 &  9:50:16.34& &   16.18 &   1.63 &   0.01 &   0.01 & &  16.21 &   1.61 &   0.01 &   0.01  \\ 
094 & 5:34:43.18 & 10:01:59.92& &   16.29 &   1.76 &   0.02 &   0.01 & &  16.28 &   1.75 &   0.01 &   0.01  \\ 
095 & 5:35:24.21 &  9:55:14.94& &   16.31 &   1.64 &   0.01 &   0.01 & &  16.35 &   1.61 &   0.01 &   0.01  \\ 
096 & 5:35:11.17 &  9:57:20.9 & &   16.37 &   1.65 &   0.02 &   0.01 & &  ---   &  ---   &  ---   &  ---    \\ 
097 & 5:35:09.26 &  9:45:59.5 & &   16.39 &   1.62 &   0.01 &   0.01 & &  ---   &  ---   &  ---   &  ---    \\ 
098 & 5:36:31.53 &  9:45:01.41& &   16.41 &   1.72 &   0.02 &   0.01 & &  16.40 &   1.72 &   0.01 &   0.01  \\ 
099 & 5:34:45.59 & 10:05:48.64& &   16.45 &   1.68 &   0.02 &   0.01 & &  16.42 &   1.72 &   0.01 &   0.01  \\ 
100 & 5:35:00.11 &  9:46:14.34& &   16.45 &   1.63 &   0.01 &   0.01 & &  16.43 &   1.65 &   0.01 &   0.01  \\ 
\hline
\end{tabular}
\end{table*}

\setcounter{table}{0}
\begin{table*}
\tiny
\caption[ ]{Positions and optical photometric data for our lambda Orionis candidate members.}
\begin{tabular}{lllcccccccccc}
\hline
LOri&   R.A      & DEC         &\, & Ic    &(R-I)c&$\delta$R&$\delta$I&\,& Ic   &(R-I)c&$\delta$R&$\delta$I \\ 
\cline{2-3} \cline{5-8} \cline{10-13}
CFHT&  \multicolumn{2}{c}{(2000.0)}  & & \multicolumn{4}{c}{shallow exp.} & & \multicolumn{4}{c}{deep exp.}\\
\#  & (h:m:s)    & ($^\circ$ ' '') &       &      &         &         &      &      &         &          \\&                       \\
\hline
101 & 5:36:30.50 & 10:08:55.52& &   16.57 &   1.60 &   0.01 &   0.01 & &  16.48 &   1.66 &   0.01 &   0.01  \\ 
102 & 5:35:22.04 &  9:52:52.19& &   16.51 &   1.75 &   0.02 &   0.01 & &  16.50 &   1.74 &   0.01 &   0.01  \\ 
103 & 5:35:22.56 &  9:45:02.12& &   16.58 &   1.72 &   0.02 &   0.01 & &  16.55 &   1.75 &   0.01 &   0.01  \\ 
104 & 5:35:07.05 &  9:54:01.1 & &   16.71 &   1.77 &   0.03 &   0.01 & &  ---   &  ---   &  ---   &  ---    \\ 
105 & 5:34:17.57 &  9:52:30.14& &   16.78 &   1.78 &   0.02 &   0.01 & &  16.75 &   1.83 &   0.01 &   0.01  \\ 
106 & 5:35:28.80 &  9:54:09.92& &   16.79 &   1.71 &   0.02 &   0.01 & &  16.76 &   1.72 &   0.01 &   0.01  \\ 
107 & 5:35:55.17 &  9:52:20.35& &   16.85 &   1.99 &   0.04 &   0.01 & &  16.78 &   2.07 &   0.01 &   0.01  \\ 
108 & 5:35:26.03 & 10:08:10.05& &   16.75 &   1.88 &   0.03 &   0.01 & &  16.80 &   1.84 &   0.01 &   0.01  \\ 
109 & 5:34:08.54 &  9:50:43.57& &   16.87 &   1.78 &   0.01 &   0.01 & &  16.81 &   1.86 &   0.01 &   0.01  \\ 
110 & 5:35:32.63 &  9:52:48.87& &   16.83 &   1.65 &   0.02 &   0.01 & &  16.82 &   1.72 &   0.01 &   0.01  \\ 
111 & 5:34:19.55 & 10:09:08.83& &   16.91 &   1.92 &   0.03 &   0.01 & &  16.86 &   2.02 &   0.01 &   0.01  \\ 
112 & 5:34:33.57 &  9:43:55.59& &   16.94 &   1.79 &   0.03 &   0.01 & &  16.87 &   1.85 &   0.01 &   0.01  \\ 
113 & 5:33:47.92 & 10:01:39.62& &   17.01 &   1.70 &   0.03 &   0.01 & &  16.99 &   1.72 &   0.01 &   0.01  \\ 
114 & 5:36:18.10 &  9:52:25.53& &   17.08 &   1.92 &   0.04 &   0.01 & &  17.06 &   1.93 &   0.01 &   0.01  \\ 
115 & 5:34:46.32 & 10:02:32.00& &   ---   &  ---   &  ---   &  ---   & &  17.08 &   1.72 &   0.01 &   0.01  \\ 
116 & 5:35:12.07 & 10:01:04.75& &   17.25 &   1.84 &   0.04 &   0.01 & &  17.17 &   1.88 &   0.01 &   0.01  \\ 
117 & 5:35:07.95 & 10:00:06.25& &   17.26 &   1.93 &   0.06 &   0.01 & &  17.21 &   2.03 &   0.01 &   0.01  \\ 
118 & 5:35:24.42 &  9:53:51.73& &   17.25 &   1.84 &   0.04 &   0.01 & &  17.23 &   1.87 &   0.01 &   0.01  \\ 
119 & 5:34:19.49 &  9:42:22.73& &   17.35 &   1.76 &   0.04 &   0.01 & &  17.30 &   1.81 &   0.01 &   0.01  \\ 
120 & 5:34:46.20 &  9:55:36.90& &   17.36 &   1.83 &   0.04 &   0.01 & &  17.34 &   1.89 &   0.01 &   0.01  \\ 
121 & 5:34:34.37 &  9:42:16.61& &   ---   &  ---   &  ---   &  ---   & &  17.37 &   1.75 &   0.01 &   0.01  \\ 
122 & 5:34:35.43 &  9:51:18.71& &   17.45 &   1.84 &   0.05 &   0.01 & &  17.38 &   1.93 &   0.01 &   0.01  \\ 
123 & 5:34:20.47 & 10:05:22.4 & &   17.42 &   2.11 &   0.06 &   0.01 & &  ---   &  ---   &  ---   &  ---    \\ 
124 & 5:34:14.24 &  9:48:26.97& &   ---   &  ---   &  ---   &  ---   & &  17.45 &   1.85 &   0.01 &   0.01  \\ 
125 & 5:34:14.24 &  9:48:26.9 & &   17.51 &   1.78 &   0.04 &   0.01 & &  ---   &  ---   &  ---   &  ---    \\ 
126 & 5:35:39.88 &  9:53:23.64& &   17.56 &   1.94 &   0.06 &   0.01 & &  17.52 &   2.00 &   0.01 &   0.01  \\ 
127 & 5:34:11.26 &  9:51:30.70& &   17.53 &   2.34 &   0.10 &   0.01 & &  ---   &  ---   &  ---   &  ---    \\ 
128 & 5:35:06.29 &  9:58:02.82& &   17.62 &   1.95 &   0.07 &   0.01 & &  17.58 &   1.95 &   0.01 &   0.01  \\ 
129 & 5:36:09.84 &  9:42:37.44& &   17.60 &   1.86 &   0.05 &   0.01 & &  17.59 &   1.92 &   0.01 &   0.01  \\ 
130 & 5:34:56.54 &  9:42:32.45& &   17.64 &   1.81 &   0.05 &   0.01 & &  17.63 &   1.81 &   0.01 &   0.01  \\ 
131 & 5:36:07.02 &  9:52:51.65& &   17.61 &   2.08 &   0.07 &   0.01 & &  17.78 &   2.01 &   0.01 &   0.01  \\ 
132 & 5:34:29.18 &  9:47:07.70& &   17.86 &   2.14 &   0.09 &   0.01 & &  17.82 &   2.17 &   0.01 &   0.01  \\ 
133 & 5:36:13.95 & 10:08:10.69& &   ---   &  ---   &  ---   &  ---   & &  17.83 &   1.85 &   0.01 &   0.01  \\ 
134 & 5:35:22.88 &  9:55:06.65& &   17.92 &   2.17 &   0.11 &   0.01 & &  17.90 &   2.01 &   0.01 &   0.01  \\ 
135 & 5:35:09.35 &  9:52:43.94& &   17.93 &   1.88 &   0.08 &   0.01 & &  17.90 &   2.01 &   0.01 &   0.01  \\ 
136 & 5:34:38.31 &  9:58:13.1 & &   17.92 &   2.14 &   0.12 &   0.01 & &  ---   &  ---   &  ---   &  ---    \\ 
137 & 5:36:30.91 & 10:05:13.5 & &   17.96 &   1.93 &   0.08 &   0.09 & &  ---   &  ---   &  ---   &  ---    \\ 
138 & 5:33:43.44 &  9:45:22.81& &   18.03 &   2.07 &   0.11 &   0.01 & &  17.96 &   2.05 &   0.01 &   0.01  \\ 
139 & 5:35:44.34 & 10:05:54.11& &   ---   &  ---   &  ---   &  ---   & &  18.16 &   1.88 &   0.01 &   0.01  \\ 
140 & 5:34:19.29 &  9:48:28.02& &   18.25 &   2.08 &   0.12 &   0.02 & &  18.21 &   2.13 &   0.01 &   0.01  \\ 
141 & 5:35:38.08 &  9:51:05.22& &   18.83 &   2.07 &   0.22 &   0.03 & &  18.25 &   2.19 &   0.01 &   0.01  \\ 
142 & 5:34:17.00 & 10:06:16.42& &   18.36 &   2.06 &   0.14 &   0.02 & &  18.27 &   2.07 &   0.01 &   0.01  \\ 
143 & 5:35:00.94 &  9:58:21.59& &   ---   &  ---   &  ---   &  ---   & &  18.30 &   2.02 &   0.01 &   0.01  \\ 
144 & 5:34:20.07 &  9:59:27.3 & &   18.30 &   1.94 &   0.11 &   0.11 & &  ---   &  ---   &  ---   &  ---    \\ 
145 & 5:36:32.83 &  9:56:01.10& &   ---   &  ---   &  ---   &  ---   & &  18.37 &   2.28 &   0.02 &   0.01  \\ 
146 & 5:35:00.15 &  9:52:40.7 & &   18.60 &   2.28 &   0.26 &   0.02 & &  ---   &  ---   &  ---   &  ---    \\ 
147 & 5:35:06.30 &  9:46:54.24& &   ---   &  ---   &  ---   &  ---   & &  18.60 &   1.94 &   0.01 &   0.01  \\ 
148 & 5:36:29.00 &  9:43:21.45& &   18.58 &   2.17 &   0.20 &   0.03 & &  18.62 &   2.15 &   0.02 &   0.01  \\ 
149 & 5:33:42.75 & 10:05:33.02& &   ---   &  ---   &  ---   &  ---   & &  18.95 &   2.12 &   0.02 &   0.01  \\ 
150 & 5:35:07.48 &  9:49:33.64& &   ---   &  ---   &  ---   &  ---   & &  19.00 &   2.29 &   0.03 &   0.01  \\ 
\hline
\end{tabular}
\end{table*}

\setcounter{table}{0}
\begin{table*}
\tiny
\caption[ ]{Positions and optical photometric data for our lambda Orionis candidate members.}
\begin{tabular}{lllcccccccccc}
\hline
LOri&   R.A      & DEC         &\, & Ic    &(R-I)c&$\delta$R&$\delta$I&\,& Ic   &(R-I)c&$\delta$R&$\delta$I \\ 
\cline{2-3} \cline{5-8} \cline{10-13}
CFHT&  \multicolumn{2}{c}{(2000.0)}  & & \multicolumn{4}{c}{shallow exp.} & & \multicolumn{4}{c}{deep exp.}\\
\#  & (h:m:s)    & ($^\circ$ ' '') &       &      &         &         &      &      &         &          \\&                       \\
\hline
151 & 5:35:58.65 &  9:48:54.74& &   ---   &  ---   &  ---   &  ---   & &  19.00 &   1.98 &   0.02 &   0.01  \\ 
152 & 5:34:11.26 &  9:44:26.70& &   ---   &  ---   &  ---   &  ---   & &  19.05 &   2.38 &   0.04 &   0.01  \\ 
153 & 5:36:18.20 &  9:57:40.97& &   ---   &  ---   &  ---   &  ---   & &  19.17 &   2.13 &   0.03 &   0.01  \\ 
154 & 5:34:19.78 &  9:54:20.84& &   ---   &  ---   &  ---   &  ---   & &  19.31 &   2.48 &   0.05 &   0.01  \\ 
155 & 5:36:25.07 & 10:01:54.32& &   ---   &  ---   &  ---   &  ---   & &  19.36 &   2.51 &   0.06 &   0.01  \\ 
156 & 5:34:36.28 &  9:55:32.18& &   ---   &  ---   &  ---   &  ---   & &  19.59 &   2.46 &   0.06 &   0.01  \\ 
157 & 5:34:11.26 &  9:55:36.84& &   ---   &  ---   &  ---   &  ---   & &  19.63 &   2.46 &   0.06 &   0.01  \\ 
158 & 5:34:20.40 & 10:03:47.61& &   ---   &  ---   &  ---   &  ---   & &  19.67 &   2.40 &   0.05 &   0.01  \\ 
159 & 5:36:33.18 & 10:00:34.29& &   ---   &  ---   &  ---   &  ---   & &  20.01 &   2.24 &   0.06 &   0.01  \\ 
160 & 5:34:11.27 &  9:45:10.72& &   ---   &  ---   &  ---   &  ---   & &  20.29 &   2.53 &   0.13 &   0.02  \\ 
161 & 5:35:54.10 &  9:43:36.11& &   ---   &  ---   &  ---   &  ---   & &  20.34 &   2.75 &   0.19 &   0.01  \\ 
162 & 5:35:04.44 &  9:57:33.64& &   ---   &  ---   &  ---   &  ---   & &  20.42 &   2.80 &   0.51 &   0.02  \\ 
163 & 5:35:18.36 &  9:56:52.78& &   ---   &  ---   &  ---   &  ---   & &  20.42 &   2.54 &   0.24 &   0.02  \\ 
164 & 5:34:11.24 &  9:52:49.35& &   ---   &  ---   &  ---   &  ---   & &  20.44 &   2.67 &   0.17 &   0.01  \\ 
165 & 5:35:11.57 &  9:53:00.56& &   ---   &  ---   &  ---   &  ---   & &  20.73 &   2.39 &   0.22 &   0.02  \\ 
166 & 5:34:00.35 &  9:54:22.45& &   ---   &  ---   &  ---   &  ---   & &  20.75 &   2.58 &   0.18 &   0.02  \\ 
167 & 5:35:14.19 &  9:54:07.52& &   ---   &  ---   &  ---   &  ---   & &  20.90 &   2.96 &   0.64 &   0.02  \\ 
168 & 5:34:50.32 &  9:45:16.92& &   ---   &  ---   &  ---   &  ---   & &  21.54 &   2.61 &   0.62 &   0.04  \\ 
169 & 5:34:59.08 &  9:58:55.60& &   ---   &  ---   &  ---   &  ---   & &  21.88 &   2.95 &   1.10 &   0.05  \\ 
170 & 5:35:36.88 &  9:44:24.41& &   ---   &  ---   &  ---   &  ---   & &  22.06 &   3.35 &   2.61 &   0.07  \\ 
\hline
\end{tabular}
$\,$\\
$\delta$$R$  and $\delta$$I$ correspond to the internal errors
as computed with IRAF. For other sources of the photmetric errors,
see the text.
\end{table*}


\setcounter{table}{1}
\begin{table*}
\tiny
\caption[ ]{Positions and IR photometry from 2MASS for our lambda Orionis candidate members.}
\begin{tabular}{lcccccccc}
\hline
LOri&   Ic   &  dist  &     RA  (2000.0)  DEC     &  J       errorJ &  H   errorH  &  K  error K  & Selection$\dagger$ &Mem \\  
CFHT&        &        &                           &                 &              &              &                    &    \\
\hline
001 & 12.52 &   0.49 &   05:33:47.21  +09:55:38.5 &  11.297   0.022 &  10.595   0.022 &  10.426   0.021 & Y,Y,Y,Y,Y, - & Mem  \\ %
002 & 12.64 &   0.06 &   05:36:10.36  +10:08:54.8 &  11.230   0.024 &  10.329   0.023 &  10.088   0.019 & N,N,Y,Y,N, - & NM-  \\ %
003 & 12.65 &   0.65 &   05:35:55.43  +09:56:30.9 &  11.416   0.023 &  10.725   0.022 &  10.524   0.023 & Y,Y,Y,Y,Y, - & Mem  \\ %
004 & 12.65 &   0.63 &   05:35:47.58  +09:45:50.0 &  11.359   0.022 &  10.780   0.023 &  10.548   0.021 & Y,Y,Y,Y,Y, - & Mem  \\ %
005 & 12.67 &   0.45 &   05:33:53.66  +09:43:08.4 &  11.378   0.022 &  10.549   0.022 &  10.354   0.023 & N,N,Y,Y,N, - & NM-  \\ %
006 & 12.75 &   0.87 &   05:35:16.23  +09:55:18.5 &  11.542   0.026 &  10.859   0.026 &  10.648   0.021 & Y,Y,Y,Y,Y, - & Mem  \\ %
007 & 12.78 &   0.51 &   05:34:29.53  +09:48:58.3 &  11.698   0.027 &  11.101   0.024 &  10.895   0.030 & Y,Y,Y,Y,Y, - & Mem  \\ %
008 & 12.79 &   0.16 &   05:35:57.98  +09:54:32.8 &  11.548   0.029 &  10.859   0.023 &  10.651   0.024 & Y,Y,Y,Y,Y, - & Mem  \\ %
009 & 12.95 &   0.26 &   05:34:46.35  +10:06:35.8 &  11.843   0.024 &  11.109   0.024 &  10.923   0.023 & Y,Y,Y,Y,Y, - & Mem  \\ %
010 & 12.96 &   0.11 &   05:34:32.97  +10:08:41.2 &  11.880   0.026 &  11.219   0.026 &  11.041   0.023 & Y,Y,Y,Y,Y, - & Mem  \\ %
011 & 13.01 &   0.51 &   05:34:44.66  +09:53:58.0 &  11.604   0.026 &  10.784   0.024 &  10.554   0.024 & N,N,Y,Y,N, - & NM-  \\ %
012 & 13.03 &   0.28 &   05:35:05.96  +09:52:08.0 &  11.816   0.026 &  10.971   0.024 &  10.795   0.023 & N,Y,Y,Y,N, - & NM-  \\ %
013 & 13.03 &   0.27 &   05:33:56.34  +09:53:56.9 &  11.656   0.022 &  10.918   0.022 &  10.719   0.023 & Y,Y,Y,Y,Y, - & Mem  \\ %
014 & 13.03 &   0.13 &   05:36:19.04  +10:03:50.9 &  11.941   0.024 &  11.278   0.027 &  11.092   0.023 & Y,Y,Y,Y,Y, - & Mem  \\ %
015 & 13.05 &   0.45 &   05:34:21.82  +10:04:14.9 &  11.870   0.024 &  11.127   0.024 &  10.912   0.019 & Y,Y,Y,Y,Y, - & Mem  \\ %
016 & 13.18 &   0.98 &   05:35:13.46  +09:55:25.3 &  11.958   0.024 &  11.284   0.027 &  11.053   0.024 & Y,Y,Y,Y,Y, - & Mem  \\ %
017 & 13.19 &   0.35 &   05:36:20.51  +09:52:19.2 &  12.188   0.024 &  11.482   0.023 &  11.323   0.021 & Y,Y,Y,Y,Y, - & Mem  \\ %
018 & 13.26 &   0.52 &   05:36:16.61  +09:50:48.4 &  11.991   0.024 &  11.284   0.022 &  11.090   0.023 & Y,Y,Y,Y,Y, - & Mem  \\ %
019 & 13.31 &   1.56 &   05:35:13.65  +09:56:27.3 &  12.019   0.026 &  11.316   0.024 &  11.067   0.021 & Y,Y,Y,Y,Y, - & Mem  \\ %
020 & 13.31 &   0.26 &   05:34:57.57  +09:46:07.2 &  11.856   0.028 &  11.214   0.026 &  11.025   0.027 & Y,Y,Y,Y,Y, - & Mem  \\ %
021 & 13.38 &   0.34 &   05:35:06.94  +09:48:57.5 &  12.258   0.027 &  11.560   0.026 &  11.296   0.021 & Y,Y,Y,Y,Y, - & Mem  \\ %
022 & 13.38 &   0.49 &   05:35:51.34  +09:55:11.2 &  12.102   0.023 &  11.411   0.022 &  11.156   0.019 & Y,Y,Y,Y,Y, - & Mem  \\ %
023 & 13.44 &   0.68 &   05:35:57.68  +09:47:34.1 &  12.221   0.027 &  11.471   0.022 &  11.290   0.024 & Y,Y,Y,Y,Y, - & Mem  \\ %
024 & 13.45 &   0.66 &   05:34:57.12  +09:54:36.7 &  12.139   0.030 &  11.446   0.026 &  11.223   0.028 & Y,Y,Y,Y,Y, - & Mem  \\ %
025 & 13.45 &   0.31 &   05:36:20.20  +09:44:01.9 &  12.163   0.044 &  11.409   0.051 &  11.090   0.033 & N,Y,Y,Y,Y, - & Mem? \\ %
026 & 13.47 &   0.66 &   05:34:36.23  +09:53:44.2 &  12.046   0.028 &  11.324   0.024 &  11.092   0.025 & Y,Y,Y,Y,Y, - & Mem  \\ %
027 & 13.50 &   0.27 &   05:35:11.01  +10:07:36.4 &  12.378   0.026 &  11.718   0.023 &  11.503   0.021 & Y,Y,Y,Y,Y, - & Mem  \\ %
028 & 13.65 &   0.38 &   05:36:18.87  +09:51:35.1 &  12.488   0.024 &  11.872   0.022 &  11.687   0.021 & N,Y,Y,Y,Y, - & Mem? \\ %
029 & 13.69 &   0.35 &   05:35:25.37  +10:08:38.4 &  12.210   0.026 &  11.460   0.027 &  11.071   0.019 & N,N,Y,Y,Y, - & NM-  \\ %
030 & 13.74 &   0.86 &   05:35:12.54  +09:55:19.5 &  12.427   0.027 &  11.686   0.026 &  11.428   0.021 & Y,Y,Y,Y,Y, - & Mem  \\ %
031 & 13.75 &   1.30 &   05:34:49.02  +09:58:02.1 &  12.412   0.028 &  11.654   0.023 &  11.442   0.028 & Y,Y,Y,Y,Y, - & Mem  \\ %
032 & 13.80 &   0.97 &   05:36:09.32  +09:47:02.7 &  12.410   0.029 &  11.714   0.023 &  11.493   0.021 & Y,Y,Y,Y,Y, - & Mem  \\ %
033 & 13.81 &   0.20 &   05:35:34.84  +10:00:35.3 &  12.455   0.033 &  11.800   0.042 &  11.502   0.027 & Y,Y,Y,Y,Y, - & Mem  \\ %
034 & 13.97 &   0.10 &   05:35:19.92  +10:02:36.5 &  12.442   0.026 &  11.639   0.026 &  11.184   0.023 & N,N,Y,Y,Y, - & NM-  \\ %
035 & 13.97 &   0.42 &   05:35:15.14  +10:01:06.8 &  12.546   0.024 &  11.842   0.027 &  11.609   0.019 & Y,Y,Y,Y,Y, - & Mem  \\ %
036 & 13.98 &   0.87 &   05:34:39.29  +10:01:28.7 &  12.576   0.024 &  11.936   0.023 &  11.706   0.021 & Y,Y,Y,Y,Y, - & Mem  \\ %
037 & 13.99 &   1.14 &   05:34:35.61  +09:59:43.3 &  12.459   0.024 &  11.727   0.026 &  11.492   0.021 & Y,Y,Y,Y,Y, - & Mem  \\ %
038 & 14.01 &   0.73 &   05:33:49.93  +09:50:36.8 &  12.684   0.030 &  11.954   0.029 & 011.752    --   & Y,Y,Y,Y,Y, - & Mem  \\ %
039 & 14.02 &   0.61 &   05:35:57.09  +09:46:52.6 &  12.755   0.030 &  12.004   0.023 &  11.775   0.023 & Y,Y,Y,Y,Y, - & Mem  \\ %
040 & 14.06 &   0.25 &   05:35:39.48  +09:50:32.8 &  12.553   0.024 &  11.877   0.022 &  11.594   0.024 & Y,Y,Y,Y,Y, - & Mem  \\ %
041 & 14.10 &   0.45 &   05:35:30.45  +09:50:34.1 &  12.500   0.024 &  11.856   0.023 &  11.587   0.027 & Y,Y,Y,Y,Y, - & Mem  \\ %
042 & 14.14 &   0.33 &   05:36:07.11  +10:09:47.7 &  12.813   0.027 &  12.099   0.026 &  11.853   0.023 & Y,Y,Y,Y,Y, - & Mem  \\ %
043 & 14.16 &   1.83 &   05:35:02.74  +09:56:47.6 &  12.707   0.024 &  12.021   0.026 &  11.741   0.024 & Y,Y,Y,Y,Y, - & Mem  \\ %
044 & 14.17 &   0.17 &   05:34:08.39  +09:51:25.3 &  12.924   0.024 &  12.318   0.024 &  12.065   0.023 & Y,Y,Y,Y,Y, - & Mem  \\ %
045 & 14.23 &   1.42 &   05:35:07.42  +09:58:22.4 &  12.768   0.023 &  12.102   0.026 &  11.844   0.023 & Y,Y,Y,Y,Y, - & Mem  \\ %
046 & 14.36 &   0.33 &   05:34:26.08  +09:51:49.4 &  13.033   0.023 &  12.478   0.026 &  12.252   0.026 & N,Y,Y,Y,Y, - & Mem? \\ %
047 & 14.38 &   0.61 &   05:35:55.67  +09:50:53.3 &  12.732   0.026 &  12.097   0.031 &  11.827   0.026 & Y,Y,Y,Y,Y, - & Mem  \\ %
048 & 14.41 &   0.49 &   05:35:12.56  +09:53:11.1 &  12.887   0.027 &  12.196   0.029 &  11.932   0.026 & Y,Y,Y,Y,Y, - & Mem  \\ %
049 & 14.50 &   0.27 &   05:35:01.00  +09:49:36.1 &  13.173   0.027 &  12.592   0.029 &  12.253   0.023 & Y,Y,Y,Y,Y, - & Mem  \\ %
050 & 14.54 &   0.81 &   05:34:56.40  +09:55:04.6 &  12.877   0.027 &  12.236   0.027 &  11.955   0.031 & Y,Y,Y,Y,Y, - & Mem  \\ %
\hline
\end{tabular}
$\,$ \\
$\dagger$ Selection criteria:
(1) [K,J-K]; (2) [I,I-K]; (3) [I-J,H-K]; (4) [I-J,I-K];(5) [J-H,H-K];
(6) Spectral type. 
\end{table*}

\setcounter{table}{1}
\begin{table*}
\tiny
\caption[ ]{Positions and IR photometry from 2MASS for our lambda Orionis candidate members.}
\begin{tabular}{lcccccccc}
\hline
LOri&   Ic   &  dist  &     RA  (2000.0)  DEC     &  J       errorJ &  H   errorH  &  K  error K  & Selection$\dagger$ &Mem \\  
CFHT&        &        &                           &                 &              &              &                    &    \\
\hline
051 & 14.60 &   0.41 &   05:36:12.14  +10:00:57.0 &  13.266   0.024 &  12.559   0.022 &  12.285   0.021 & Y,Y,Y,Y,Y, - & Mem  \\ %
052 & 14.63 &   0.55 &   05:34:11.78  +09:57:03.4 &  13.117   0.023 &  12.454   0.024 &  12.192   0.019 & Y,Y,Y,Y,Y, - & Mem  \\ %
053 & 14.72 &   0.34 &   05:34:36.73  +09:52:58.3 &  13.173   0.032 &  12.521   0.023 &  12.278   0.027 & Y,Y,Y,Y,Y, - & Mem  \\ %
054 & 14.73 &   0.74 &   05:35:52.55  +09:48:31.1 &  13.189   0.024 &  12.509   0.022 &  12.271   0.027 & Y,Y,Y,Y,Y, - & Mem  \\ %
055 & 14.76 &   0.30 &   05:35:21.43  +09:49:56.6 &  13.184   0.026 &  12.477   0.026 &  12.253   0.026 & Y,Y,Y,Y,Y, - & Mem  \\ %
056 & 14.87 &   0.44 &   05:34:58.37  +09:53:47.1 &  13.211   0.029 &  12.567   0.026 &  12.267   0.029 & Y,Y,Y,Y,Y, - & Mem  \\ %
057 & 15.04 &   0.74 &   05:35:11.32  +10:00:50.2 &  13.412   0.024 &  12.773   0.023 &  12.487   0.030 & Y,Y,Y,Y,Y, - & Mem  \\ %
058 & 15.06 &   0.71 &   05:36:10.36  +10:00:18.4 &  13.521   0.024 &  12.935   0.022 &  12.643   0.027 & Y,Y,Y,Y,Y, - & Mem  \\ %
059 & 15.10 &   0.28 &   05:34:23.55  +09:43:43.4 &  13.574   0.026 &  12.884   0.026 &  12.682   0.032 & Y,Y,Y,Y,Y, - & Mem  \\ %
060 & 15.14 &   0.38 &   05:35:20.00  +09:49:06.2 &  13.598   0.030 &  12.961   0.030 &  12.663   0.029 & Y,Y,Y,Y,Y, - & Mem  \\ %
061 & 15.15 &   0.15 &   05:35:18.18  +09:52:24.2 &  13.533   0.023 &  12.833   0.026 &  12.525   0.027 & Y,Y,Y,Y,Y, - & Mem  \\ %
062 & 15.16 &   0.35 &   05:35:15.33  +09:48:37.0 &  13.634   0.029 &  13.005   0.030 &  12.725   0.027 & Y,Y,Y,Y,Y, - & Mem  \\ %
063 & 15.34 &   0.60 &   05:35:19.14  +09:54:42.4 &  13.756   0.029 &  13.066   0.029 &  12.663   0.030 & N,Y,Y,Y,Y, - & Mem? \\ %
064 & 15.34 &   0.49 &   05:35:52.01  +09:50:29.4 &  13.782   0.026 &  13.098   0.025 &  12.846   0.029 & Y,Y,Y,Y,Y, - & Mem  \\ %
065 & 15.37 &   1.27 &   05:35:17.92  +09:56:57.2 &  13.820   0.024 &  13.123   0.029 &  12.843   0.027 & Y,Y,Y,Y,Y, - & Mem  \\ %
066 & 15.40 &   1.72 &   05:35:06.78  +09:57:27.8 &  13.506   0.024 &  12.901   0.026 &  12.654   0.029 & Y,Y,Y,Y,Y, - & Mem  \\ %
067 & 15.53 &   0.18 &   05:36:26.38  +09:45:46.6 &  14.000   0.033 &  13.356   0.027 &  13.102   0.036 & Y,Y,Y,Y,Y, - & Mem  \\ %
068 & 15.20 &   0.44 &   05:34:48.02  +09:43:26.2 &  13.521   0.027 &  12.902   0.026 &  12.628   0.027 & Y,N,Y,Y,Y, - & Mem? \\ %
069 & 15.20 &   0.21 &   05:34:43.97  +09:48:35.6 &  13.384   0.027 &  12.774   0.027 &  12.425   0.027 & Y,N,Y,Y,Y, - & Mem? \\ %
070 & 15.61 &   0.34 &   05:36:00.01  +10:05:48.8 &  14.042   0.032 &  13.405   0.029 &  13.067   0.031 & Y,Y,Y,Y,Y, - & Mem  \\ %
071 & 15.45 &   0.26 &   05:34:15.79  +10:06:54.6 &  13.749   0.030 &  13.129   0.024 &  12.839   0.031 & Y,Y,Y,Y,Y, - & Mem  \\ %
072 & 15.35 &   1.93 &   05:34:11.41  +09:42:07.9 &  13.554   0.026 &  12.944   0.032 &  12.631   0.027 & Y,N,Y,Y,Y, - & Mem? \\ %
073 & 15.28 &   0.02 &   05:34:46.82  +09:50:37.9 &  13.644   0.028 &  12.992   0.023 &  12.715   0.027 & Y,N,Y,Y,Y, - & Mem? \\ %
074 & 15.39 &   0.93 &   05:36:00.54  +09:42:39.0 &  13.663   0.026 &  13.088   0.025 &  12.720   0.024 & Y,N,Y,Y,Y, - & Mem? \\ %
075 & 15.23 &   0.61 &   05:34:55.22  +10:00:34.7 &  13.396   0.026 &  12.794   0.026 &  12.526   0.024 & Y,N,N,Y,Y, Y & Mem? \\ %
076 & 15.81 &   1.29 &   05:35:10.96  +09:57:43.8 &  14.216   0.027 &  13.527   0.027 &  13.201   0.032 & Y,Y,Y,Y,Y, - & Mem  \\ %
077 & 15.89 &   1.07 &   05:34:41.72  +09:42:42.0 &  14.031   0.027 &  13.416   0.027 &  13.109   0.035 & Y,Y,Y,Y,Y, - & Mem  \\ %
078 & 15.92 &   0.62 &   05:36:16.45  +09:50:15.9 &  14.227   0.041 &  13.593   0.053 &  13.286   0.040 & Y,Y,Y,Y,Y, - & Mem  \\ %
079 & 16.00 &   0.76 &   05:34:48.26  +09:59:53.9 &  14.221   0.032 &  13.536   0.032 &  13.338   0.039 & Y,Y,N,Y,Y, - & Mem? \\ %
080 & 16.01 &   0.34 &   05:35:30.05  +09:59:25.5 &  13.804   0.023 &  13.196   0.022 &  12.891   0.033 & Y,N,Y,Y,Y, - & Mem? \\ %
081 & 16.02 &   0.43 &   05:33:56.61  +10:06:14.9 &  14.669   0.032 &  13.692   0.032 &  13.209   0.037 & Y,Y,Y,Y,Y, Y & Mem+ \\ %
082 & 16.02 &   0.34 &   05:36:00.81  +09:52:57.1 &  14.200   0.033 &  13.570   0.025 &  13.281   0.033 & Y,Y,Y,Y,Y, Y & Mem+ \\ %
083 & 16.02 &   0.67 &   05:35:43.41  +09:54:26.8 &  14.265   0.030 &  13.638   0.035 &  13.375   0.040 & Y,Y,Y,Y,Y, - & Mem  \\ %
084 & 16.03 &   1.19 &   05:35:50.80  +10:09:45.2 &  14.077   0.024 &  13.448   0.027 &  13.188   0.034 & Y,Y,Y,Y,Y, - & Mem  \\ %
085 & 16.04 &   0.38 &   05:35:21.52  +09:53:29.2 &  14.189   0.026 &  13.622   0.037 &  13.233   0.027 & Y,Y,Y,Y,Y, - & Mem  \\ %
086 & 16.09 &   0.67 &   05:34:11.58  +09:49:15.0 &  14.482   0.032 &  13.867   0.032 &  13.503   0.040 & Y,Y,Y,Y,Y, - & Mem  \\ %
087 & 16.09 &   1.03 &   05:34:33.77  +09:55:34.2 &  14.186   0.039 &  13.601   0.030 &  13.279   0.035 & Y,Y,Y,Y,Y, Y & Mem+ \\ %
088 & 16.10 &   1.07 &   05:34:49.50  +09:58:46.8 &  14.140   0.031 &  13.543   0.037 &  13.228   0.039 & Y,Y,Y,Y,Y, - & Mem  \\ %
089 & 16.15 &   0.15 &   05:35:04.00  +10:07:26.8 &  14.380   0.032 &  13.839   0.035 &  13.512   0.039 & Y,Y,Y,Y,Y, - & Mem  \\ %
090 & 16.17 &   0.48 &   05:36:27.08  +09:51:34.7 &  14.515   0.041 &  13.881   0.023 &  13.651   0.051 & Y,Y,Y,Y,Y, - & Mem  \\ %
091 & 16.18 &   0.76 &   05:34:35.86  +09:54:26.5 &  14.184   0.032 &  13.556   0.032 &  13.289   0.031 & Y,Y,Y,Y,Y, - & Mem  \\ %
092 & 16.19 &   0.50 &   05:35:50.97  +09:51:03.5 &  14.441   0.030 &  13.841   0.038 &  13.537   0.040 & Y,Y,Y,Y,Y, - & Mem  \\ %
093 & 16.21 &   0.15 &   05:34:41.20  +09:50:16.3 &  14.462   0.030 &  13.836   0.039 &  13.604   0.052 & Y,Y,Y,Y,Y, - & Mem  \\ %
094 & 16.28 &   0.19 &   05:34:43.17  +10:01:59.8 &  14.404   0.034 &  13.802   0.030 &  13.425   0.038 & Y,Y,Y,Y,Y, - & Mem  \\ %
095 & 16.35 &   0.64 &   05:35:24.18  +09:55:15.4 &  14.564   0.033 &  13.913   0.029 &  13.613   0.048 & Y,Y,Y,Y,Y, Y & Mem+ \\ %
096 & 16.37 &   1.45 &   05:35:11.13  +09:57:19.6 &  14.627   0.038 &  13.965   0.037 &  13.638   0.047 & Y,Y,Y,Y,Y, - & Mem  \\ %
098 & 16.40 &   0.52 &   05:36:31.50  +09:45:01.7 &  14.647   0.037 &  13.985   0.045 &  13.682   0.039 & Y,Y,Y,Y,Y, Y & Mem+ \\ %
099 & 16.42 &   0.28 &   05:34:45.59  +10:05:48.9 &  14.709   0.034 &  14.074   0.035 &  13.676   0.043 & Y,Y,Y,Y,Y, - & Mem  \\ %
100 & 16.43 &   0.39 &   05:35:00.10  +09:46:14.0 &  14.768   0.044 &  14.044   0.042 &  13.821   0.044 & Y,Y,Y,Y,Y, - & Mem  \\ %
\hline
\end{tabular}
$\,$ \\
$\dagger$ Selection criteria:
(1) [K,J-K]; (2) [I,I-K]; (3) [I-J,H-K]; (4) [I-J,I-K];(5) [J-H,H-K];
(6) Spectral type. 
\end{table*}

\setcounter{table}{1}
\begin{table*}
\tiny
\caption[ ]{Positions and IR photometry from 2MASS for our lambda Orionis candidate members.}
\begin{tabular}{lcccccccc}
\hline
LOri&   Ic   &  dist  &     RA  (2000.0)  DEC     &  J       errorJ &  H   errorH  &  K  error K  & Selection$\dagger$ &Mem \\  
CFHT&        &        &                           &                 &              &              &                    &    \\
\hline
101 & 16.48 &   1.79 &   05:36:30.41  +10:08:54.3 &  15.019   0.038 &  14.372   0.044 &  14.110   0.066 & Y,N,Y,Y,Y, - & Mem? \\ %
102 & 16.50 &   0.31 &   05:35:22.02  +09:52:52.3 &  14.634   0.047 &  14.083   0.050 &  13.809   0.057 & N,Y,Y,Y,Y, - & Mem? \\ %
103 & 16.55 &   0.19 &   05:35:22.56  +09:45:01.9 &  14.643   0.029 &  14.126   0.029 &  13.833   0.055 & N,Y,Y,Y,Y, - & Mem? \\ %
104 & 16.71 &   0.52 &   05:35:07.07  +09:54:01.5 &  14.667   0.030 &  14.136   0.036 &  13.721   0.042 & Y,Y,Y,Y,Y, - & Mem  \\ %
105 & 16.75 &   0.46 &   05:34:17.58  +09:52:29.7 &  14.922   0.040 &  14.340   0.052 &  13.993   0.053 & Y,Y,Y,Y,Y, - & Mem  \\ %
106 & 16.76 &   0.46 &   05:35:28.77  +09:54:10.2 &  14.776   0.043 &  14.161   0.057 &  13.743   0.045 & Y,Y,Y,Y,Y, - & Mem  \\ %
107 & 16.78 &   0.46 &   05:35:55.19  +09:52:20.0 &  14.656   0.036 &  13.987   0.035 &  13.621   0.052 & Y,Y,Y,Y,Y, Y & Mem+ \\ %
108 & 16.80 &   0.26 &   05:35:26.04  +10:08:09.8 &  14.840   0.033 &  14.256   0.048 &  13.918   0.050 & Y,Y,Y,Y,Y, - & Mem  \\ %
109 & 16.81 &   0.04 &   05:34:08.54  +09:50:43.5 &  15.023   0.047 &  14.376   0.043 &  14.087   0.064 & Y,Y,Y,Y,Y, - & Mem  \\ %
110 & 16.82 &   0.36 &   05:35:32.61  +09:52:48.7 &  15.043   0.051 &  14.475   0.056 &  14.144   0.060 & Y,Y,Y,Y,Y, Y & Mem+ \\ %
111 & 16.86 &   0.46 &   05:34:19.57  +10:09:08.5 &  14.801   0.038 &  14.165   0.043 &  13.786   0.051 & Y,Y,Y,Y,Y, - & Mem  \\ %
112 & 16.87 &   0.57 &   05:34:33.53  +09:43:55.5 &  14.991   0.042 &  14.358   0.048 &  14.148   0.062 & N,Y,N,Y,Y, - & NM-  \\ %
113 & 16.99 &   0.11 &   05:33:47.91  +10:01:39.7 &  15.162   0.048 &  14.576   0.060 &  14.268   0.082 & Y,Y,Y,Y,Y, - & Mem  \\ %
114 & 17.06 &   0.18 &   05:36:18.11  +09:52:25.4 &  15.092   0.044 &  14.389   0.053 &  14.006   0.064 & Y,Y,Y,Y,Y, Y & Mem+ \\ %
115 & 17.08 &   0.14 &   05:34:46.32  +10:02:31.9 &  15.449   0.047 &  14.821   0.068 &  14.594   0.104 & N,N,Y,Y,Y, N & NM+  \\ %
116 & 17.17 &   0.72 &   05:35:12.03  +10:01:04.3 &  15.343   0.057 &  14.573   0.055 &  14.411   0.082 & Y,Y,N,Y,N, Y & Mem+ \\ %
118 & 17.23 &   0.31 &   05:35:24.41  +09:53:51.9 &  15.269   0.044 &  14.686   0.064 &  14.181   0.057 & Y,Y,Y,Y,N, Y & Mem+ \\ %
119 & 17.30 &   1.02 &   05:34:19.50  +09:42:23.7 &  14.760    --   &  14.262    --   &  14.548   0.106 & N,Y,N,N,N, ? & NM?  \\ %
120 & 17.34 &   0.78 &   05:34:46.21  +09:55:37.7 &  15.335   0.050 &  14.770   0.059 &  14.337   0.087 & Y,Y,Y,Y,Y, Y & Mem+ \\ %
121 & 17.37 &   0.66 &   05:34:34.34  +09:42:17.1 &  15.533   0.060 &  15.093   0.086 &  14.748   0.099 & N,N,Y,Y,N, - & NM-  \\ %
122 & 17.38 &   0.13 &   05:34:35.44  +09:51:18.6 &  15.428   0.066 &  14.852   0.060 &  14.462   0.080 & Y,Y,Y,Y,Y, - & Mem  \\ %
124 & 17.45 &   0.67 &   05:34:14.25  +09:48:26.3 &  15.661   0.073 &  15.059   0.082 &  14.778   0.112 & Y,N,Y,Y,Y, Y & Mem? \\ %
125 & 17.51 &   0.60 &   05:34:14.25  +09:48:26.3 &  15.661   0.073 &  15.059   0.082 &  14.778   0.112 & N,N,Y,Y,Y, - & NM-  \\ %
126 & 17.52 &   0.66 &   05:35:39.85  +09:53:24.1 &  15.511   0.077 &  14.911   0.096 &  14.335    --   & Y,Y,Y,Y,N, Y & Mem+ \\ %
127 & 17.53 &   1.31 &   05:34:11.18  +09:51:30.1 &  13.016   0.023 &  12.606   0.027 &  12.468   0.024 & N,N,N,i,N, - & NM-  \\ %
128 & 17.58 &   1.55 &   05:35:06.31  +09:58:01.3 &  15.624   0.077 &  15.099   0.087 &  14.769   0.109 & N,Y,Y,Y,Y, - & Mem? \\ %
129 & 17.59 &   0.56 &   05:36:09.81  +09:42:37.0 &  15.383   0.056 &  14.816   0.072 &  14.526   0.102 & N,Y,Y,Y,Y, - & Mem? \\ %
130 & 17.63 &   0.57 &   05:34:56.54  +09:42:33.0 &  15.731   0.059 &  15.265   0.092 &  14.735   0.110 & Y,Y,Y,Y,N, Y & Mem+ \\ %
131 & 17.78 &   0.47 &   05:36:06.99  +09:52:51.5 &  15.429   0.054 &  14.900   0.063 &  14.380   0.090 & Y,Y,Y,Y,N, - & Mem? \\ %
132 & 17.82 &   0.59 &   05:34:29.16  +09:47:07.2 &  15.583   0.067 &  14.962   0.078 &  14.913   0.145 & N,Y,N,Y,N, - & NM-  \\ %
133 & 17.83 &   0.55 &   05:36:13.95  +10:08:10.1 &  16.290   0.101 &  15.900   0.167 &  15.378   0.203 & N,N,Y,Y,N, N & NM+  \\ %
134 & 17.90 &   0.59 &   05:35:22.85  +09:55:07.1 &  15.543   0.057 &  14.937   0.074 &  14.666   0.107 & N,Y,N,Y,Y, N & NM+  \\ %
135 & 17.90 &   0.39 &   05:35:09.33  +09:52:44.2 &  15.671   0.072 &  15.082   0.087 &  14.908   0.138 & N,Y,N,Y,Y, Y & Mem? \\ %
136 & 17.92 &   1.69 &   05:34:38.37  +09:58:11.6 &  15.560   0.085 &  14.828   0.090 &  14.576   0.108 & Y,Y,N,Y,Y, - & Mem? \\ %
138 & 17.96 &   0.59 &   05:33:43.48  +09:45:22.9 &  15.821   0.078 &  15.204   0.083 &  14.971   0.133 & N,Y,N,Y,Y, - & NM-  \\ %
139 & 18.16 &   0.25 &   05:35:44.34  +10:05:54.4 &  16.074   0.096 &  15.205   0.098 &  14.729   0.103 & Y,Y,Y,Y,N, Y & Mem+ \\ %
140 & 18.21 &   0.56 &   05:34:19.27  +09:48:27.5 &  15.981   0.078 &  15.224   0.089 &  14.750   0.113 & Y,Y,Y,Y,Y, Y & Mem+ \\ %
141 & 18.25 &   0.31 &   05:35:38.06  +09:51:05.2 &  16.667   0.165 &  15.707   0.149 &  15.351   0.203 & Y,N,Y,Y,N, N & NM+  \\ %
142 & 18.27 &   0.57 &   05:34:17.01  +10:06:17.0 &  16.174   0.100 &  15.579   0.131 &  15.029   0.131 & Y,Y,Y,Y,N, - & Mem? \\ %
143 & 18.30 &   1.30 &   05:35:00.95  +09:58:20.3 &  15.938   0.081 &  15.446   0.110 &  14.949   0.134 & Y,Y,Y,Y,N, Y & Mem+ \\ %
146 & 18.60 &   0.22 &   05:35:00.16  +09:52:40.9 &  16.230   0.107 &  15.470   0.110 &  14.936   0.128 & Y,Y,Y,Y,Y, - & Mem  \\ %
147 & 18.60 &   0.74 &   05:35:06.26  +09:46:53.8 &  16.833   0.178 &  16.531   0.286 &  15.689   0.255 & Y,N,Y,Y,N, N & NM+  \\ %
148 & 18.62 &   1.04 &   05:36:28.96  +09:43:22.3 &  16.545   0.154 &  15.783   0.152 &  15.709   0.244 & N,N,N,Y,N, - & NM-  \\ %
150 & 19.00 &   0.80 &   05:35:07.50  +09:49:32.9 &  16.656   0.152 &  16.134   0.197 &  15.560   0.214 & Y,Y,Y,Y,N, Y & Mem+ \\ %
152 & 19.05 &   0.56 &   05:34:11.30  +09:44:26.9 &  16.773   0.173 &  16.657   0.295 &  15.870   0.285 & N,N,Y,Y,N, - & NM-  \\ %
154 & 19.31 &   0.45 &   05:34:19.81  +09:54:20.6 &  16.804   0.169 &  16.143   0.192 &  15.513   0.219 & Y,Y,Y,Y,N, Y & Mem+ \\ %
155 & 19.36 &   0.35 &   05:36:25.05  +10:01:54.4 &  16.592   0.144 &  15.742    --   &  15.043    --   & Y,Y,Y,Y,Y, Y & Mem+ \\ %
\hline
\end{tabular}
$\,$ \\
$\dagger$ Selection criteria:
(1) [K,J-K]; (2) [I,I-K]; (3) [I-J,H-K]; (4) [I-J,I-K];(5) [J-H,H-K];
(6) Spectral type. 
\end{table*}


\setcounter{table}{2}
\begin{table*}
\tiny
\caption[ ]{Spectroscopic data.}
\begin{tabular}{ccrc}
\hline
LOri-CFHT  &Sp.Type & W(Ha)  error &Member? \\ 
\#    &        &   (\AA)      &        \\&               \\  
\hline
075  &  M5.0  &   9.4   0.5 & Mem? \\ 
081  &  M5.5  &   4.2   0.4 & Mem+ \\ 
082  &  M4.5  &   8.6   0.8 & Mem+ \\ 
087  &  M4.5  &   6.7   0.4 & Mem+ \\ 
095  &  M6.0  &   7.3   0.6 & Mem+ \\ 
098  &  M5.0  &  12.9   3.1 & Mem+ \\ 
107  &  M6.0  &  11.7   1.3 & Mem+ \\ 
110  &  M5.5  &   9.1   1.6 & Mem+ \\ 
114  &  M6.5  &  10.9   0.9 & Mem+ \\ 
115  &  M5.0  &   8.5   0.4 & NM+  \\ 
116  &  M5.5  &  11.1   0.6 & Mem+ \\ 
117  &  M6.0  &  22.9   2.6 & Mem+ \\ 
118  &  M5.5  &  10.1   0.8 & Mem+ \\ 
119  &  M5.5  &   --    --  & NM?  \\ 
120  &  M5.5  &   7.4   1.1 & Mem+ \\ 
124  &  M5.5  &   8.4   0.4 & Mem? \\ 
126  &  M6.5  &  26.2   1.9 & Mem+ \\ 
130  &  M5.5  &   8.7   0.7 & Mem+ \\ 
133  &  M4.5  &   1.9   0.4 & NM+  \\ 
134  &  M5.0  &   5.9   0.5 & NM+  \\ 
135  &  M7.0  &  15.5   1.7 & Mem? \\ 
139  &  M6.0  &  19.7   1.2 & Mem+ \\ 
140  &  M7.0  &  72.8   4.2 & Mem+ \\ 
141  &  M4.5  &   4.3   0.2 & NM+  \\ 
143  &  M6.5  &  35.7   5.2 & Mem+ \\ 
147  &  M5.5  &  10.7   0.9 & NM+  \\ 
150  &  M8.0  &  15.6   1.5 & Mem+ \\ 
151  &  M5.5  &  11.6   1.0 & NM?  \\ 
154  &  M8.0  &  16.9   2.2 & Mem+ \\ 
155  &  M8.0  &  38.   15.  & Mem+ \\ 
156  &  M8.0  & 101.7   7.9 & Mem+ \\ 
161  &  M8.5  & 123.   56.  & Mem+ \\ 
165  &  M7.5$\dagger$&  16.   15.  & NM?  \\ 
\hline
\end{tabular}
$\,$ \\
$\dagger$ Due to te low S/N,  the uncertainty in the spectral type classification 
is the one subclass, compared with half for the rest of the sample.
\end{table*}


%
%

    \begin{figure*}
    \centering
    \includegraphics[width=16.8cm]{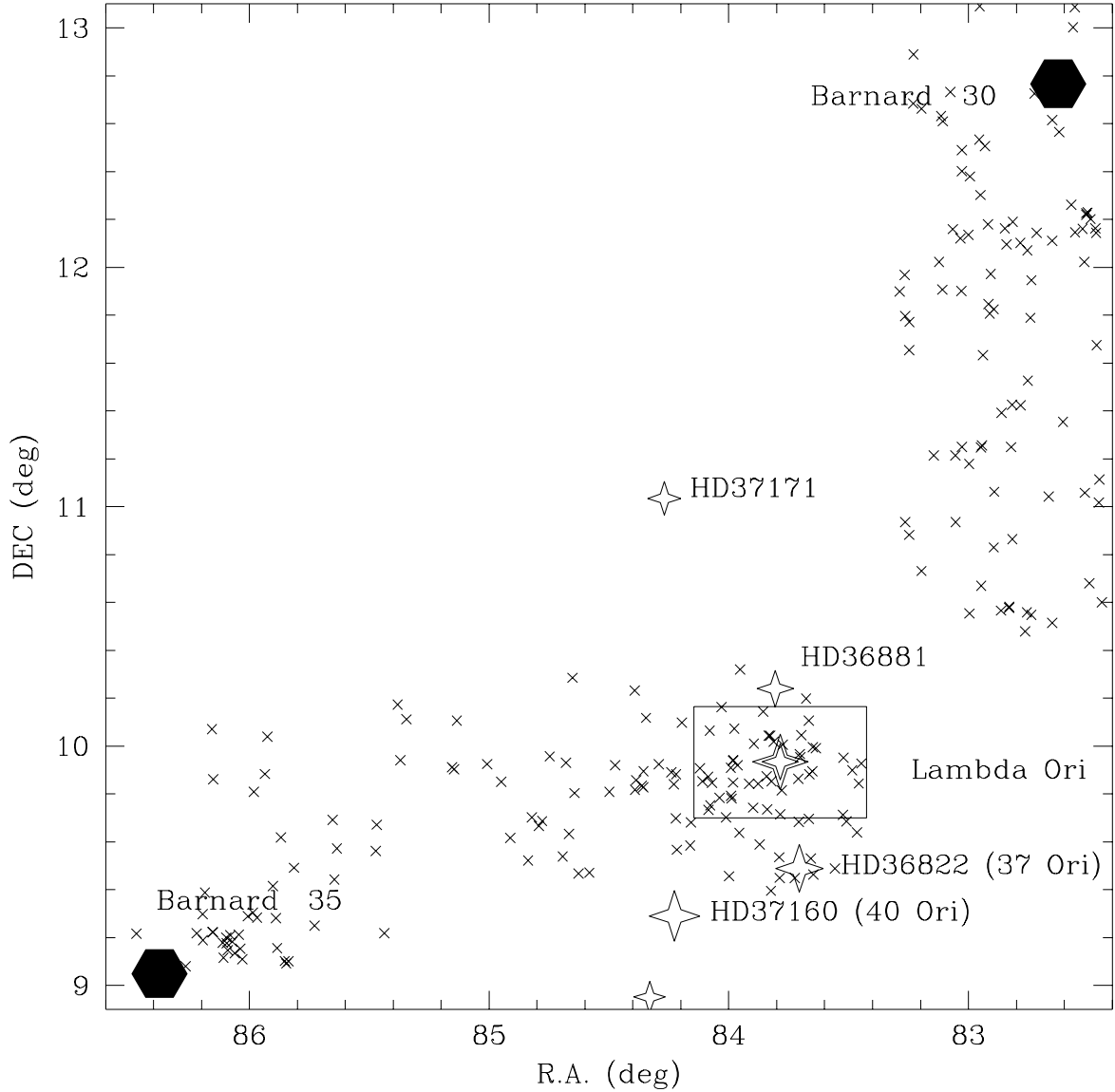}
 \caption{Location of our CFHT survey in the Cousins $R$ and $I$
filters (box). The location of the dark clouds Barnard 30 and Barnard 35
are indicated, as well as the stars listed in the Bright Star Catalog.
Pre-main sequence stars identified by Dolan \& Mathieu (1999, 2001) are
displayed as crosses. }
 \end{figure*}


\newpage

    \begin{figure*}
    \centering
    \includegraphics[width=15.8cm]{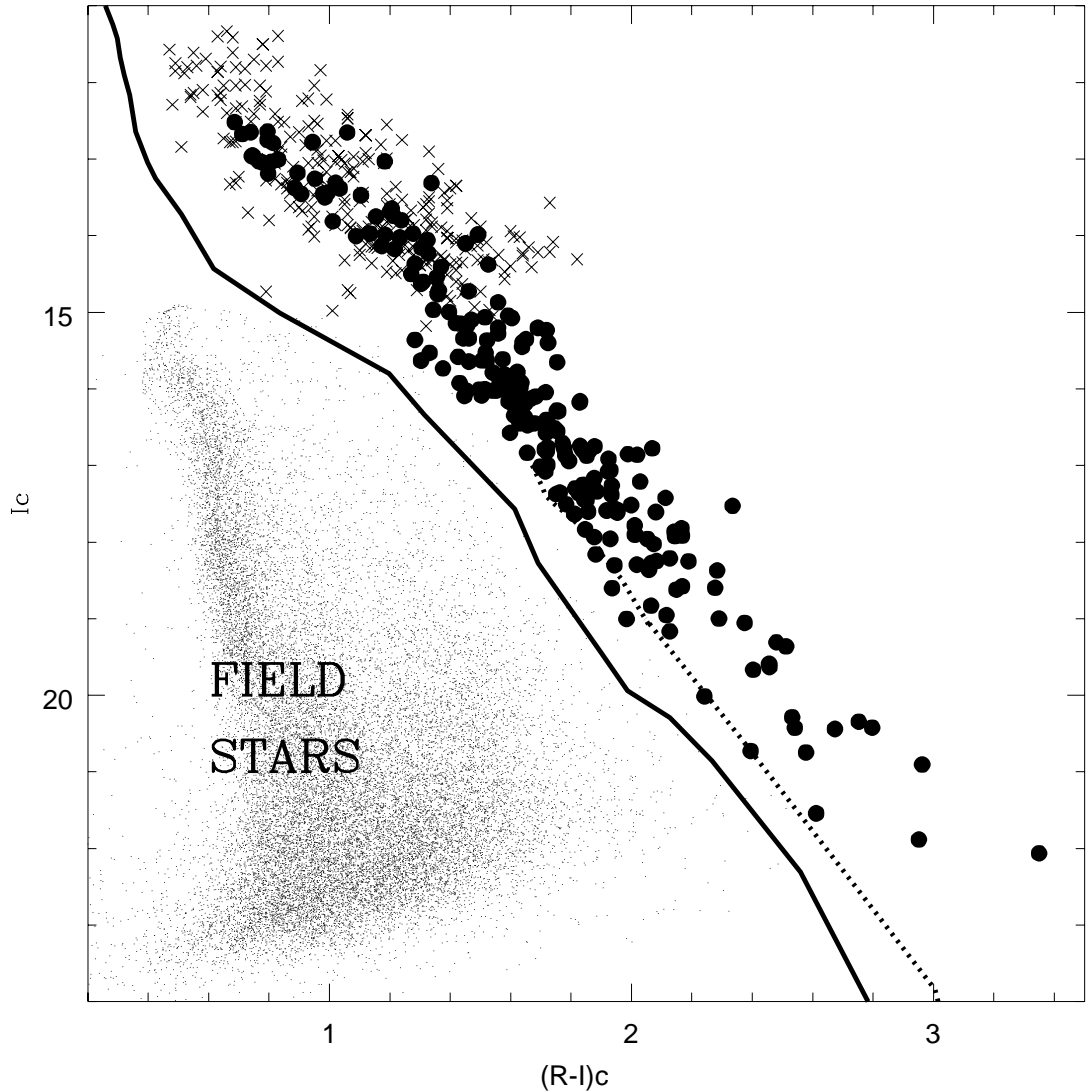}
 \caption{Optical color-magnitude diagram for the field around 
the $\lambda$ Orionis star. Field stars are displayed as dots, 
previously known cluster members, brighter than the present survey, 
appear as crosses (Dolan \& Mathieu 1999, 2001),
 whereas the new candidate 
members are included as solid circles.
The thick, dotted line corresponds to Baraffe et al. (1998) 5 Myr isochrone, 
extended toward the red end using a NextGen model. The thick, solid line 
represent an empirical ZAMS (see Barrado y Navascu\'es et al. 2001).}
 \end{figure*}


\newpage

    \begin{figure*}
    \centering
    \includegraphics[width=15.8cm]{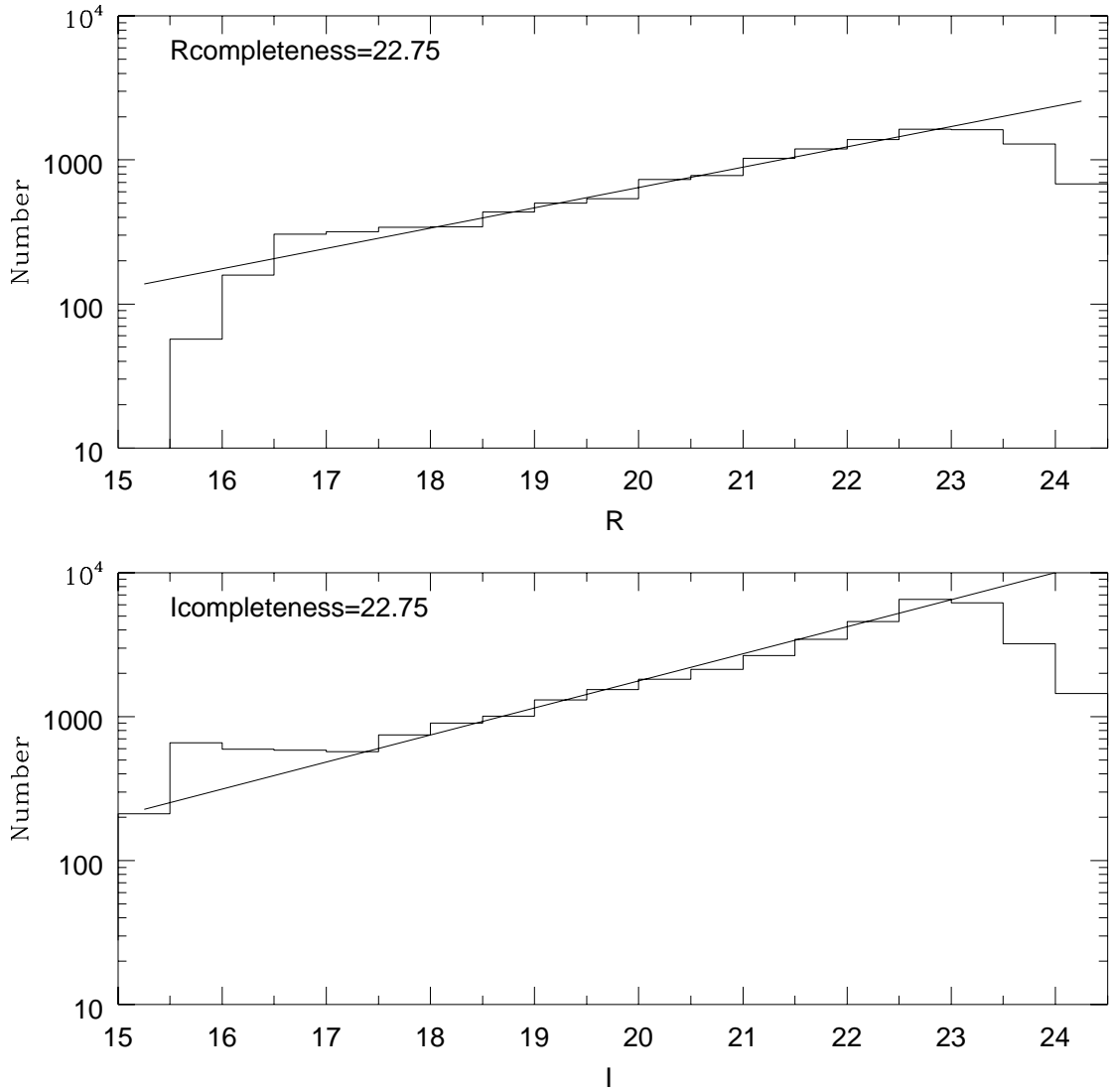}
 \caption{Completeness for the survey.
For cluster members, the completeness limits are at
I(complete)=20.2 mag.}
 \end{figure*}


\newpage

%
%

\setcounter{figure}{3}
    \begin{figure*}
    \centering
    \includegraphics[width=7.8cm]{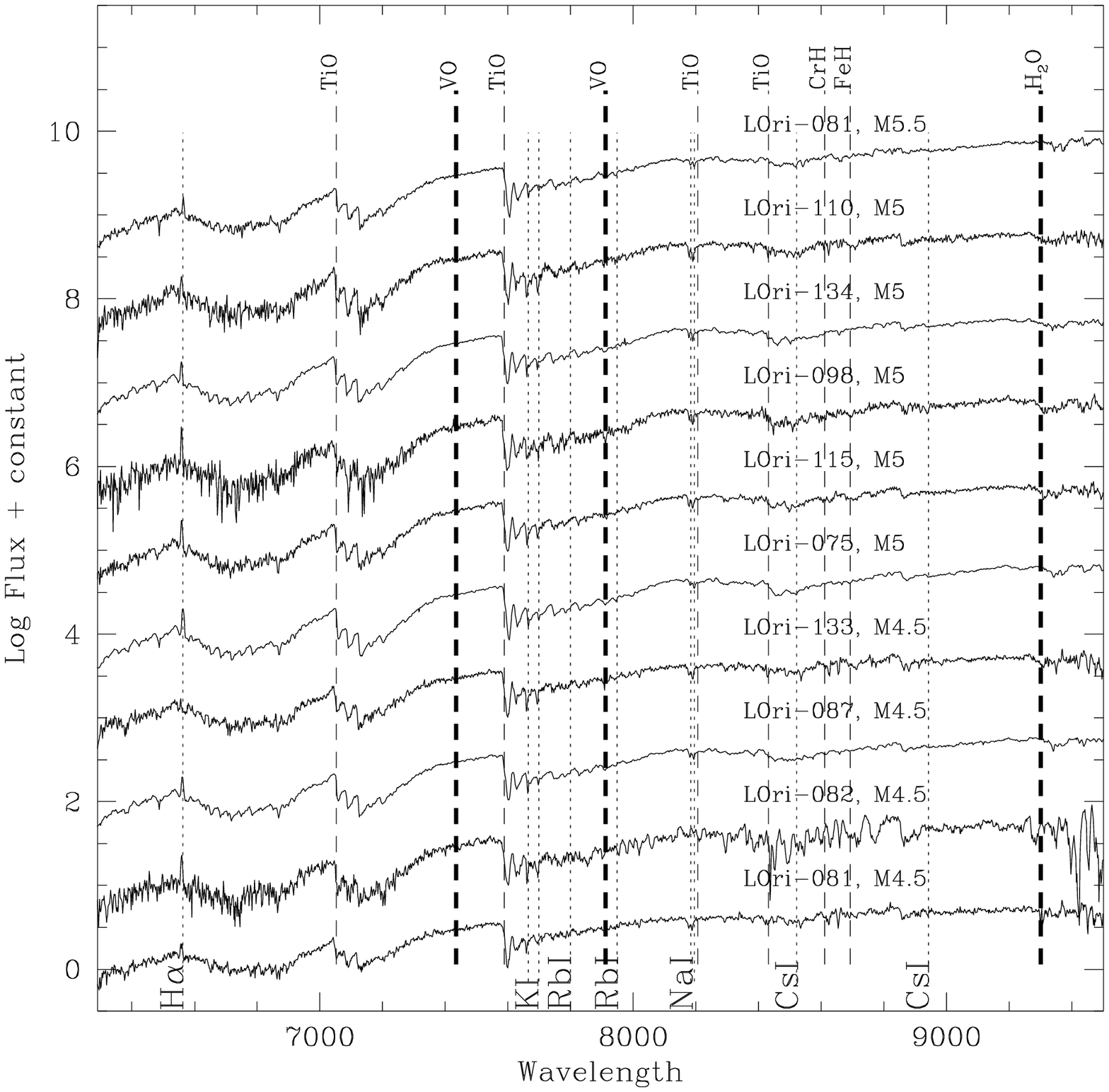}
    \includegraphics[width=7.8cm]{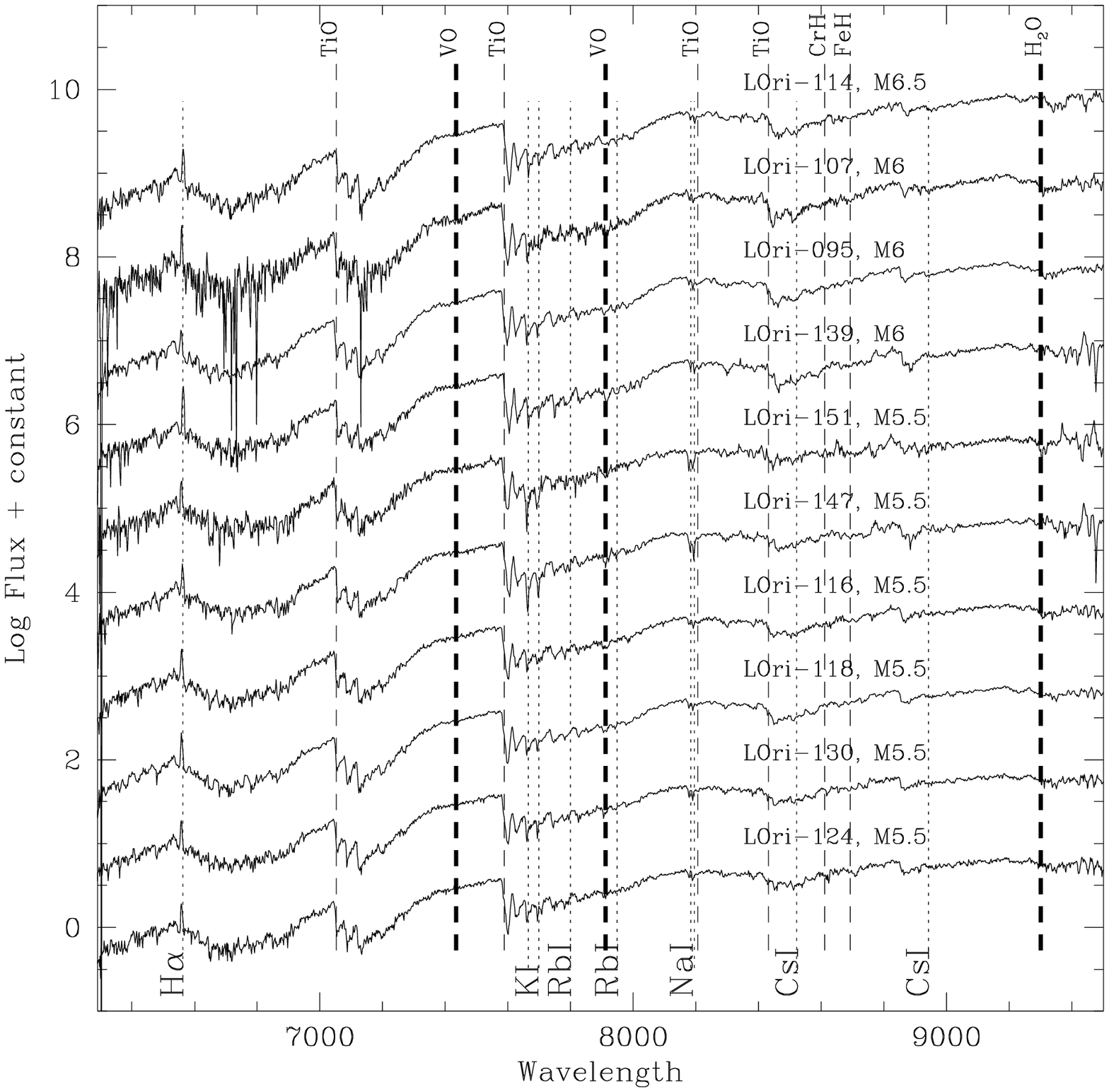}
    \includegraphics[width=7.8cm]{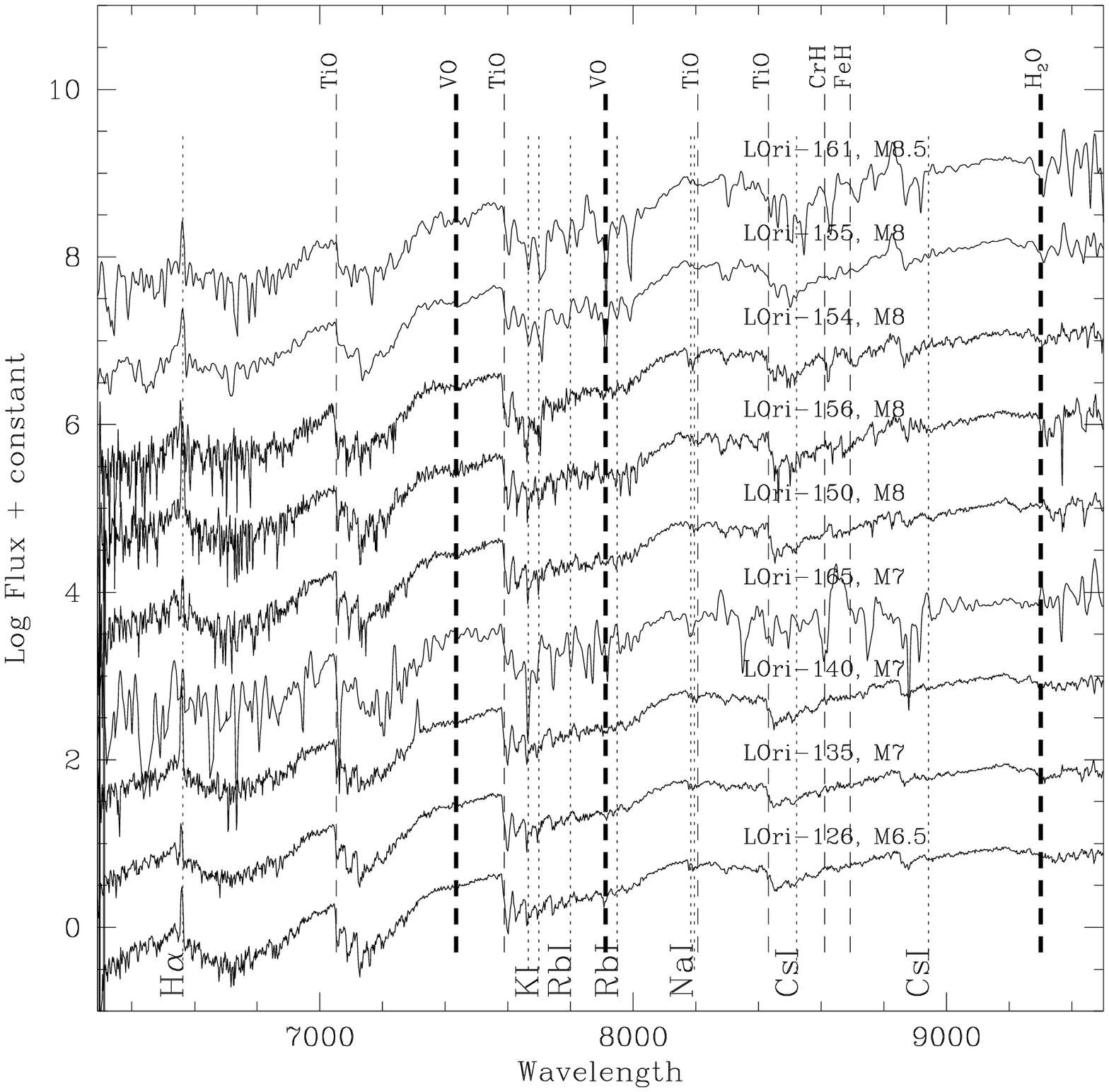}
    \includegraphics[width=7.8cm]{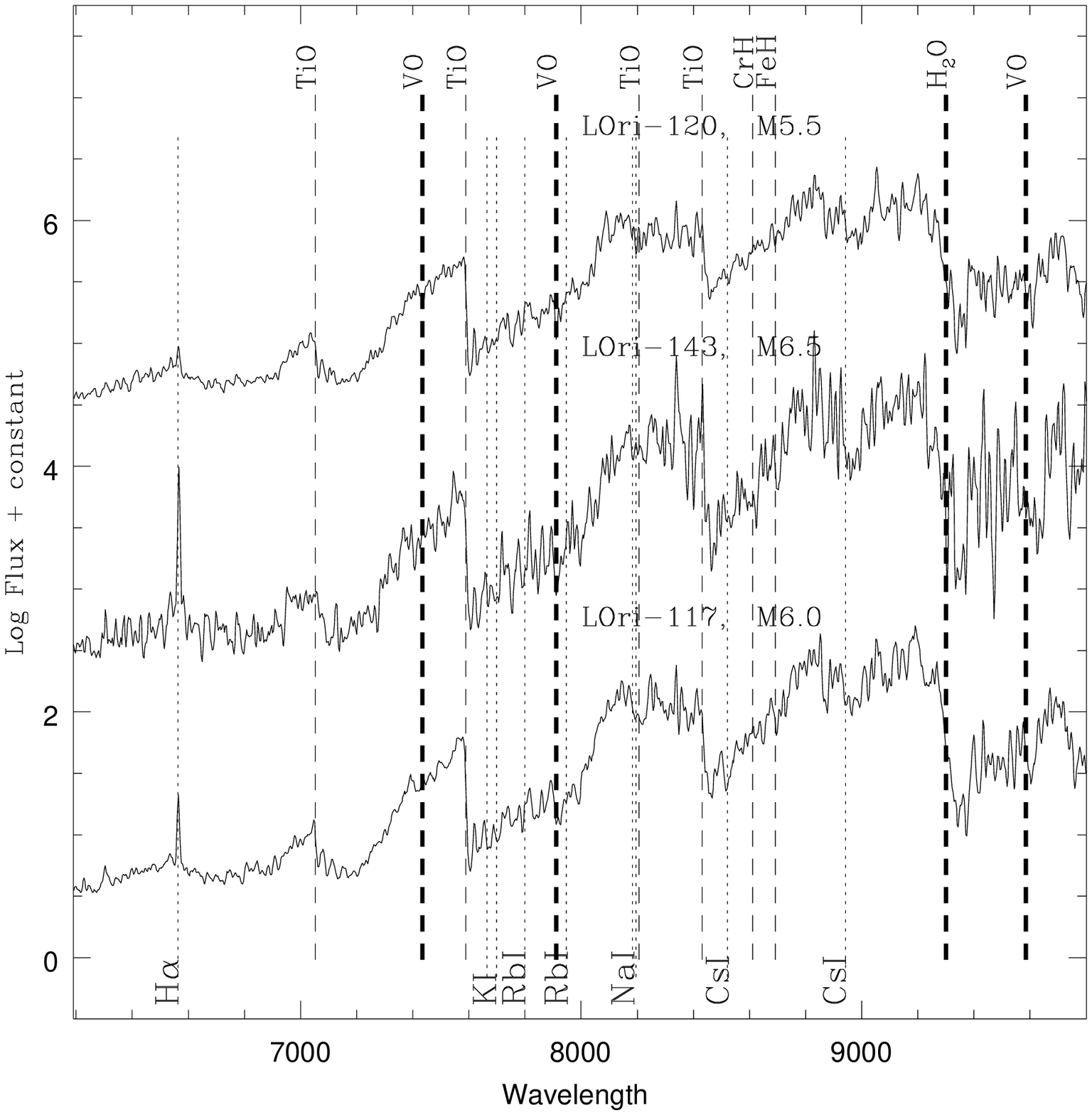}
 \caption{
{\bf a, b and c}  Keck LRIS 400 l/mm spectra.
{\bf d} Magellan B\&C 300 l/mm spectra.
}
      \end{figure*}

\newpage

%
%

\setcounter{figure}{4}
    \begin{figure*}
    \centering
    \includegraphics[width=9.8cm]{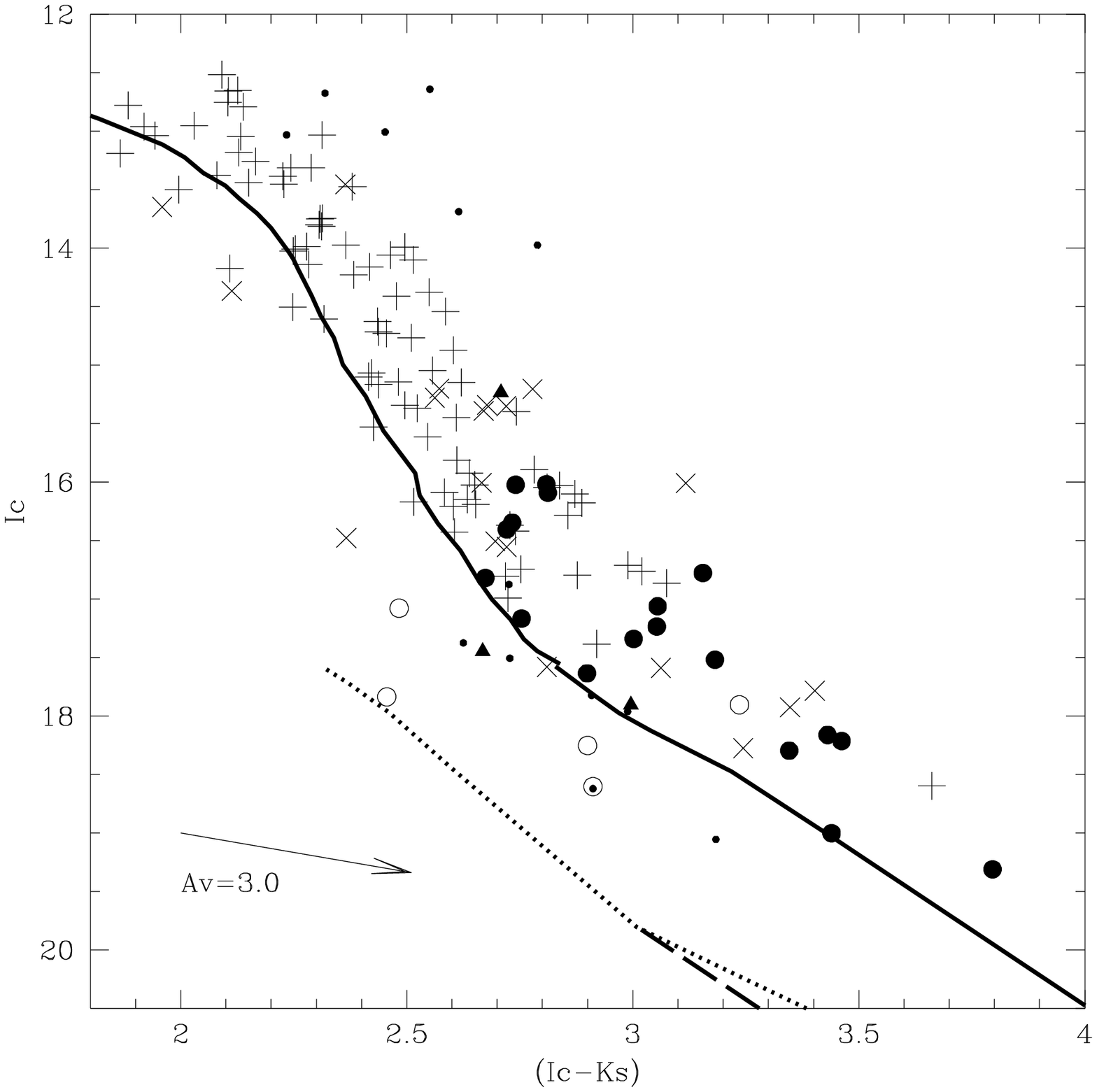}
    \includegraphics[width=9.8cm]{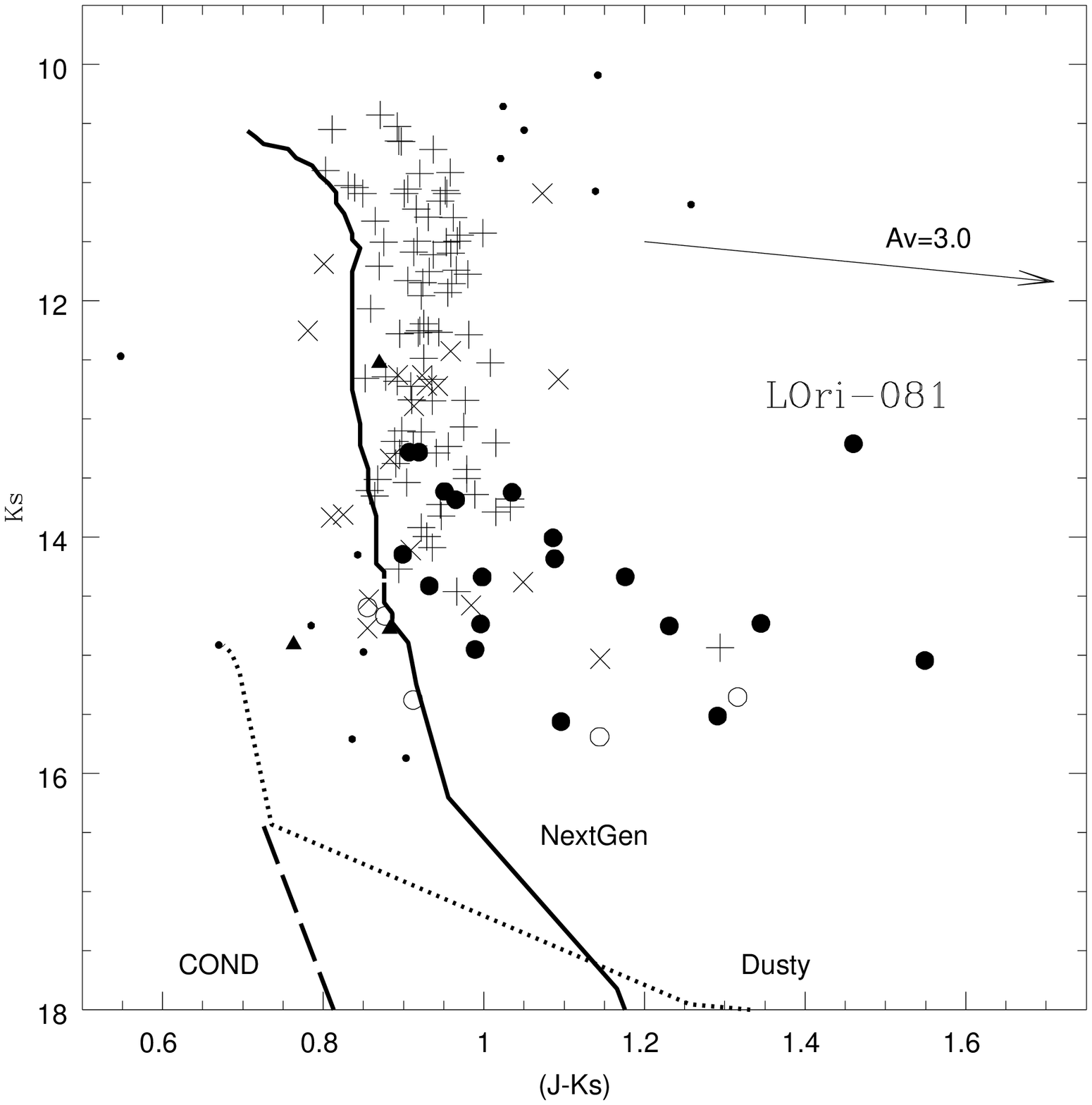}
 \caption{
Color-magnitude diagram for Lambda Orionis 
candidate members. Those members without spectroscopy are 
displayed as:
dots for ''NM-'', crosses for ''Mem?'', plus symbols  for ''mem''.
Candidates with  low resolution spectroscopy appear as:
open circles for ''NM+'', open triangles for ``NM?'', 
solid triangles for ''Mem?'',
and solid circles for ''Mem+''. 
``NextGen'''', ``Dusty'' and
``COND'' models --5Myr isochrones from Baraffe et al. (1998, 2002)
and Chabrier et al. (2000)-- are included as solid, 
dotted and  dashed lines, respectively.}
 \end{figure*}


\newpage

\setcounter{figure}{5}
    \begin{figure*}
    \centering
    \includegraphics[width=7.8cm]{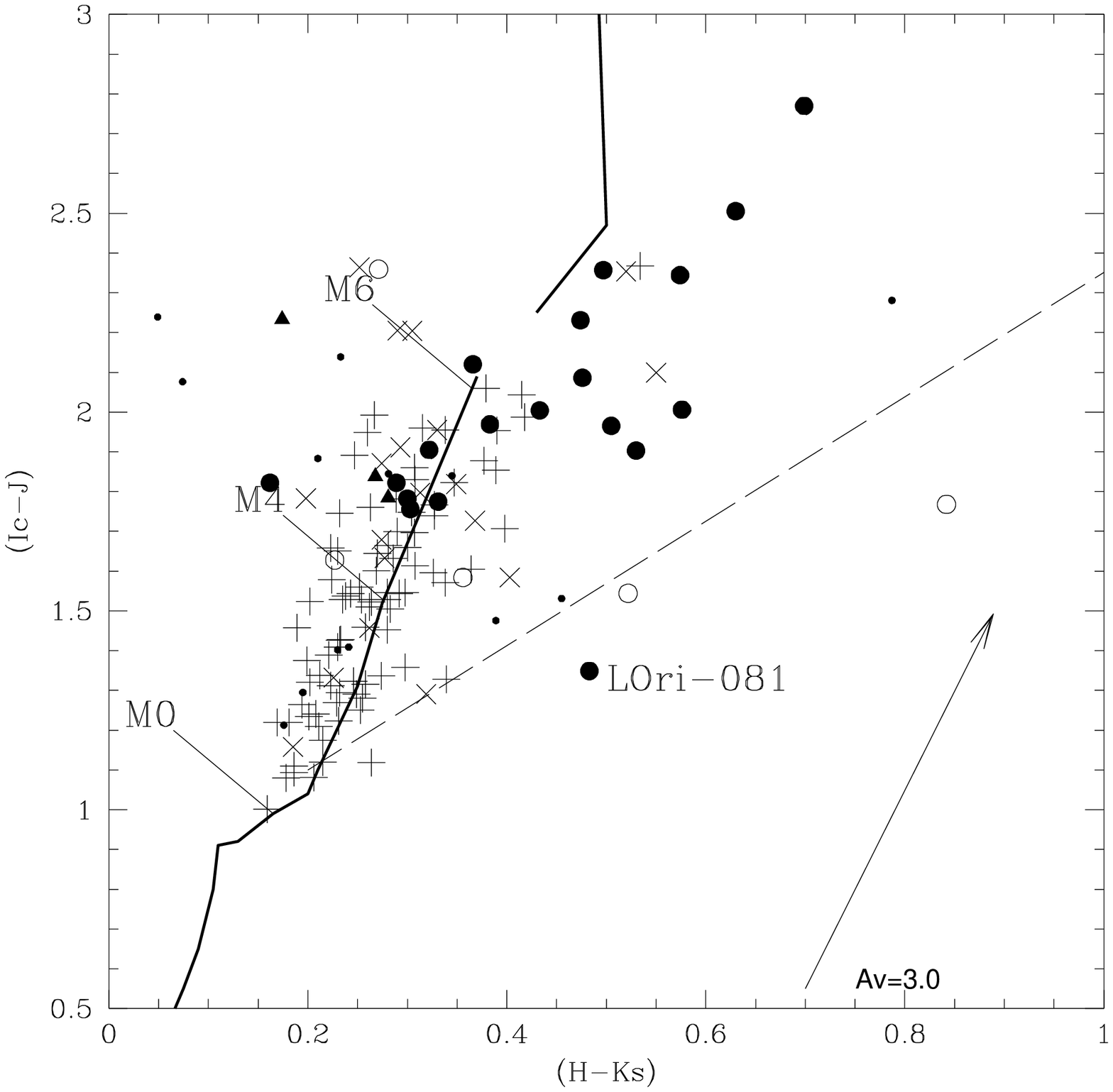}
    \includegraphics[width=7.8cm]{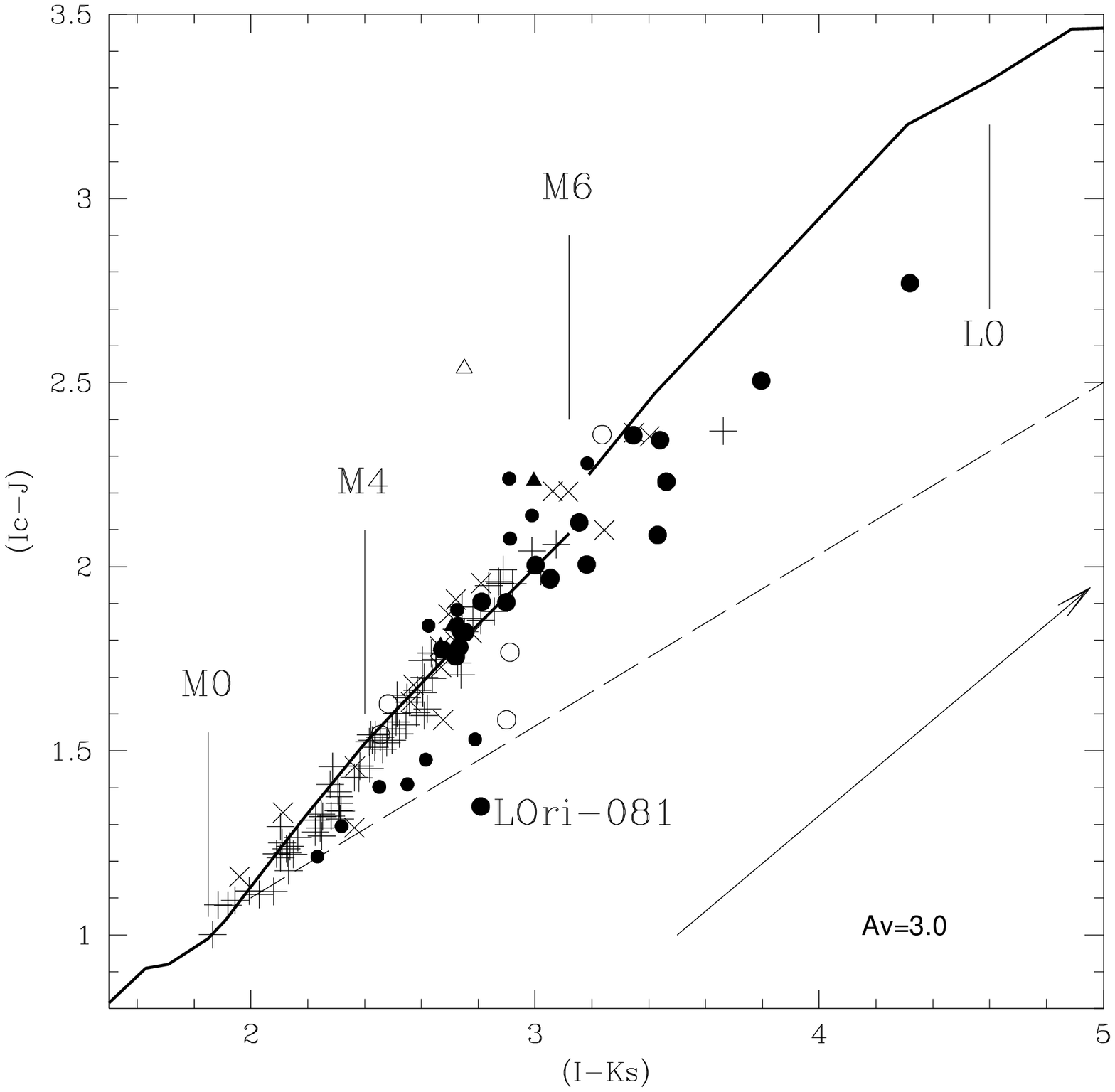}
    \includegraphics[width=7.8cm]{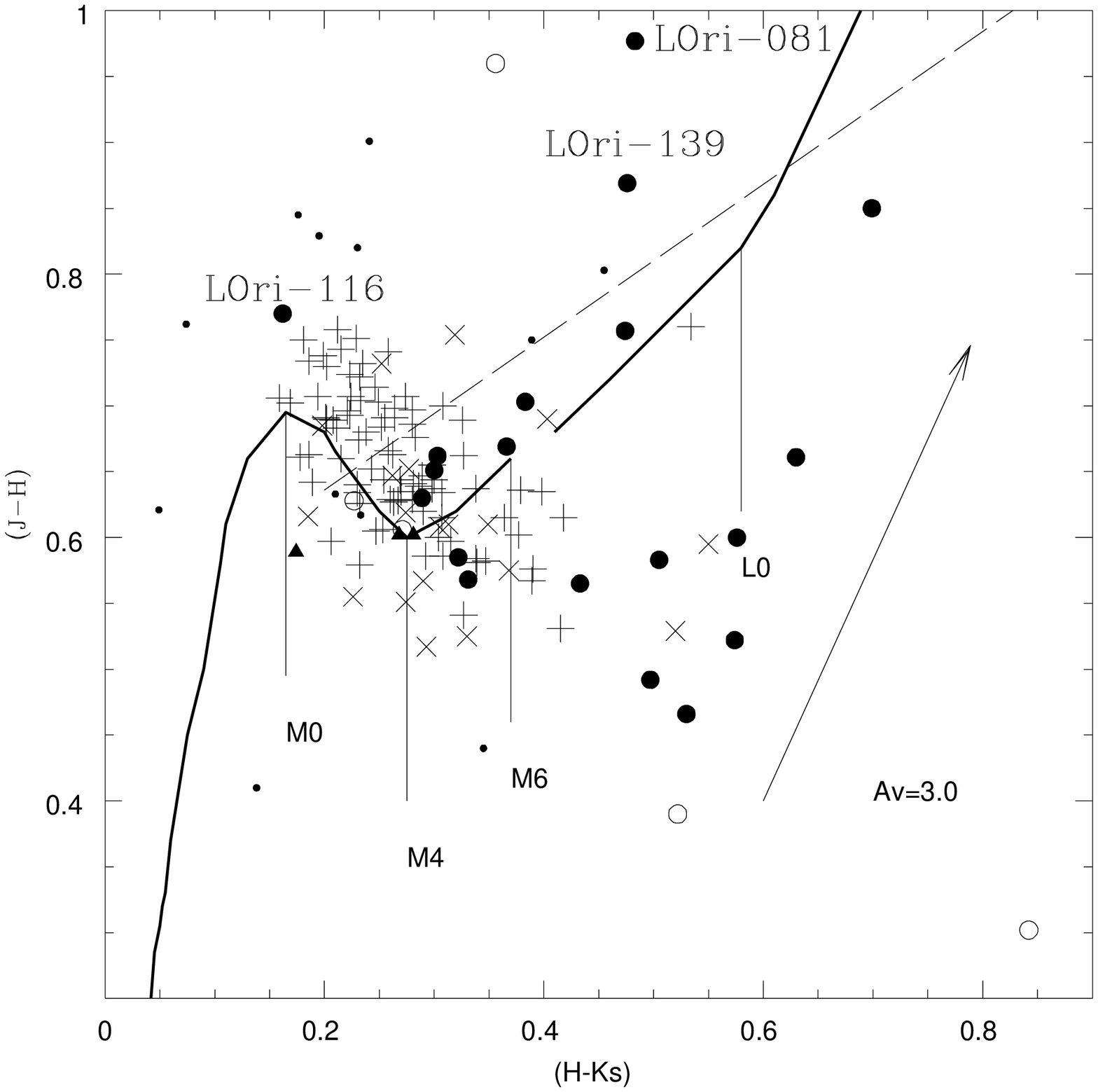}
 \caption{
Optical-Infrared color-color diagram for Lambda Orionis 
candidate members. Those members without spectroscopy are 
displayed as:
dots for ''NM-'', crosses for ''Mem?'', plus symbols for ''mem''.
Candidates with  low resolution spectroscopy appear as:
open circles for ''NM+'', open triangles for ``NM?'', 
 solid triangles for ''Mem?'',
and solid circles for ''Mem+''.
The thick-solid and dashed lines correspond to the locii of the main
 sequence stars (from Bessell \& Brett 1988; Kirkpatrick et al$.$
 2000; Leggett et al. 2001) 
and CTT stars (Meyer et al$.$ 1997;  Barrado y Navascu\'es et al. 2003),
 respectively.  
}
 \end{figure*}


\newpage

%
%

\setcounter{figure}{6}
    \begin{figure*}
    \centering
    \includegraphics[width=9.8cm]{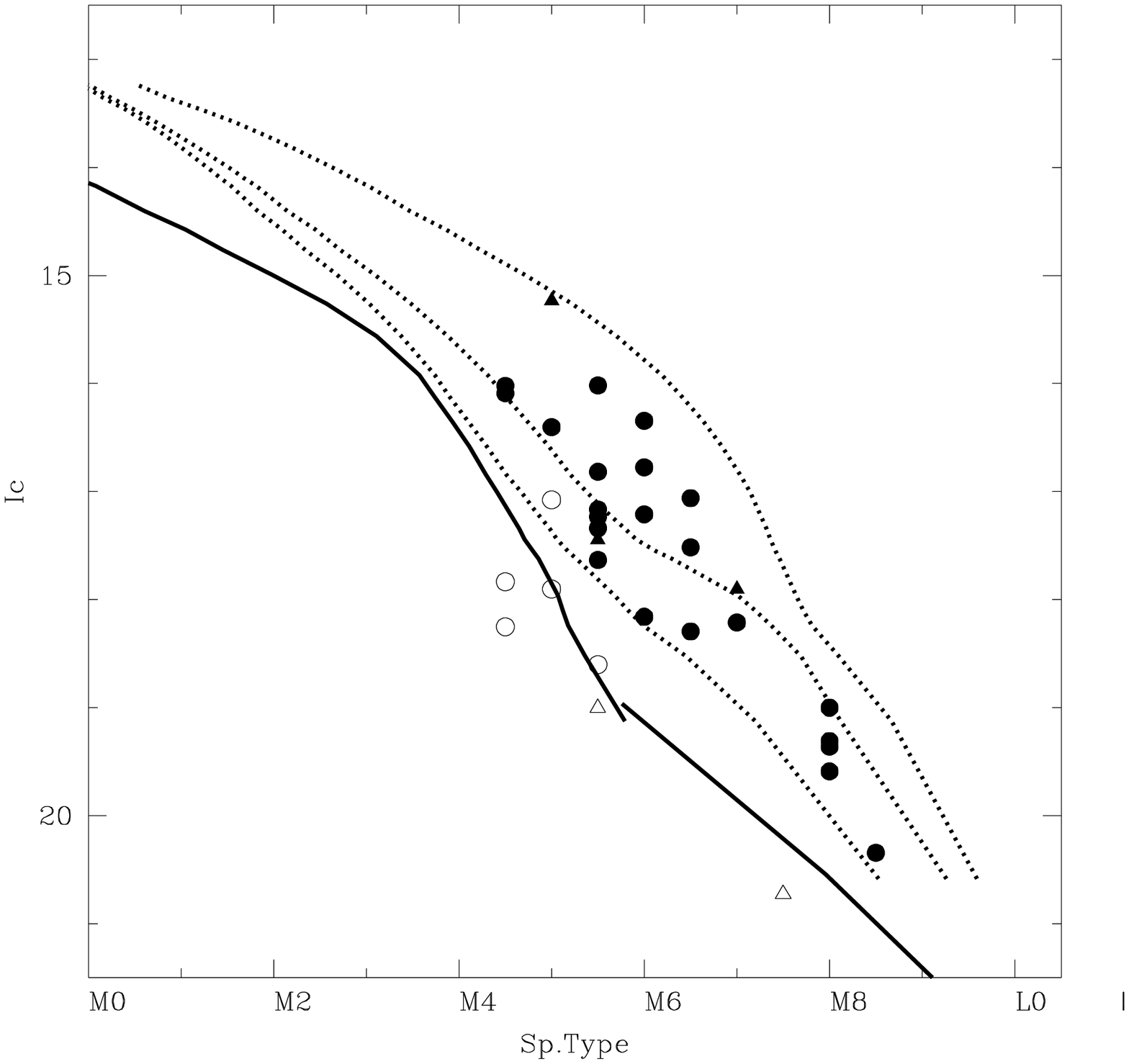}
    \includegraphics[width=9.8cm]{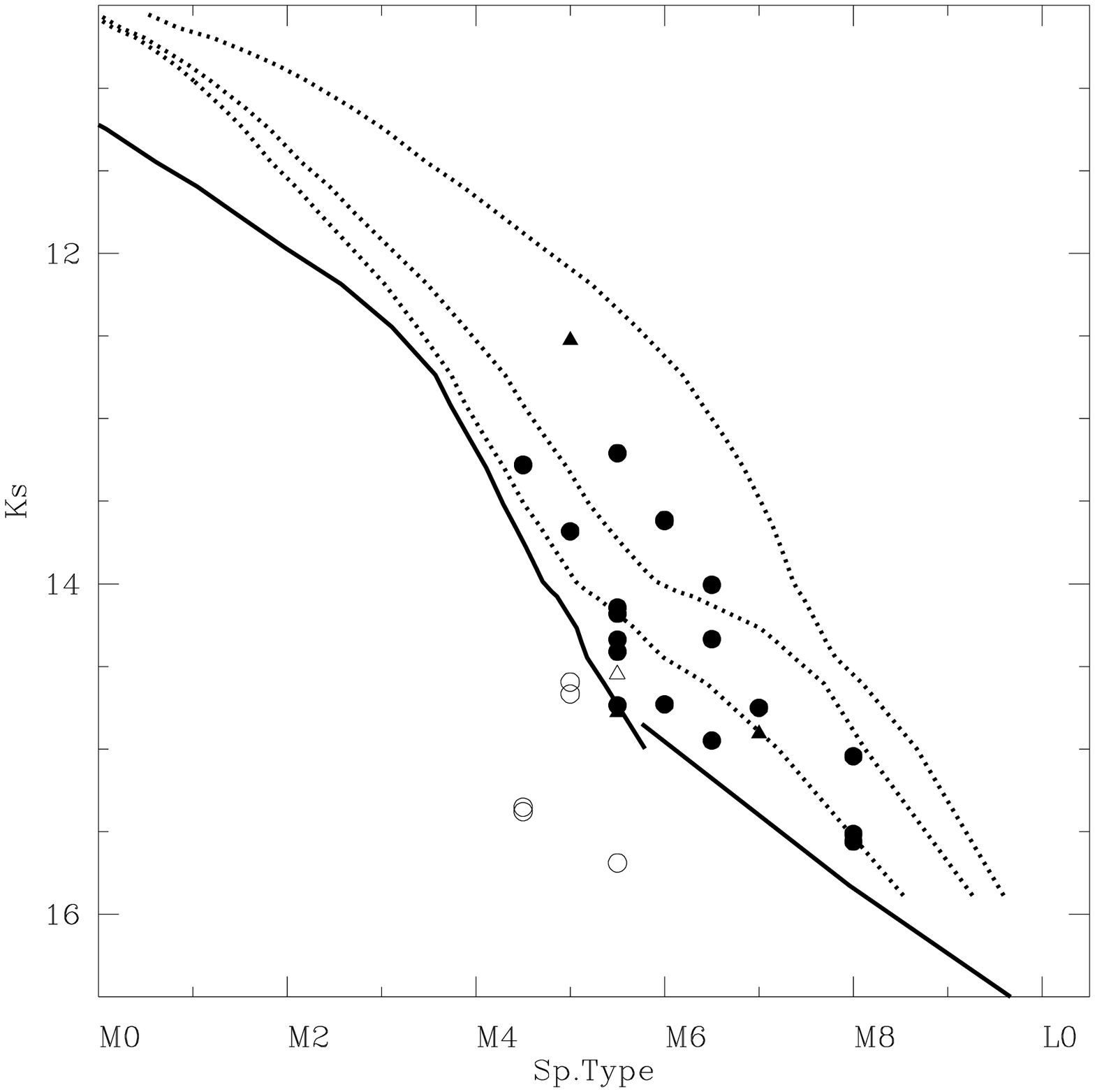}
 \caption{
{\bf a} $Ic$ magnitude versus the spectral type. Symbols as
in figure 5. The isochrones --5 Myr-- correspond to models 
by Baraffe et al. (1998), after applying different temperature scales
(Basri et al. 2000, solid line; Luhman 1999, dotted lines for
 different gravities).
{\bf b}  $Ks$ magnitude versus the spectral type. Symbols as
in  figure 5. The isochrones correspond to models 
by Baraffe et al. (1998), after applying different temperature scales
(Basri et al. 2000, solid line; Luhman 1999, dotted lines for
 different gravities).
}
 \end{figure*}

\newpage

%
%

\setcounter{figure}{7}
    \begin{figure*}
    \centering
    \includegraphics[width=9.8cm]{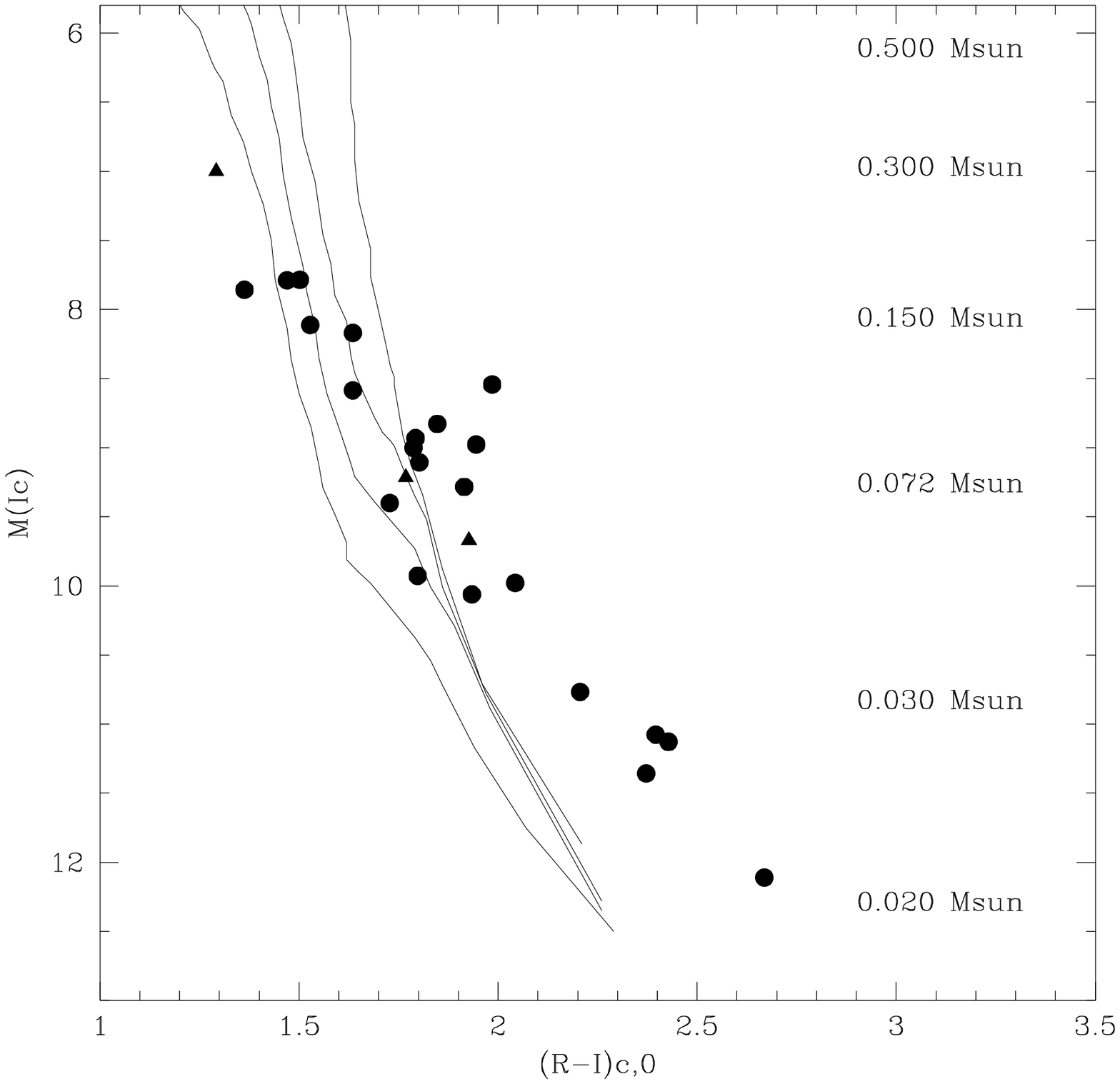}
    \includegraphics[width=9.8cm]{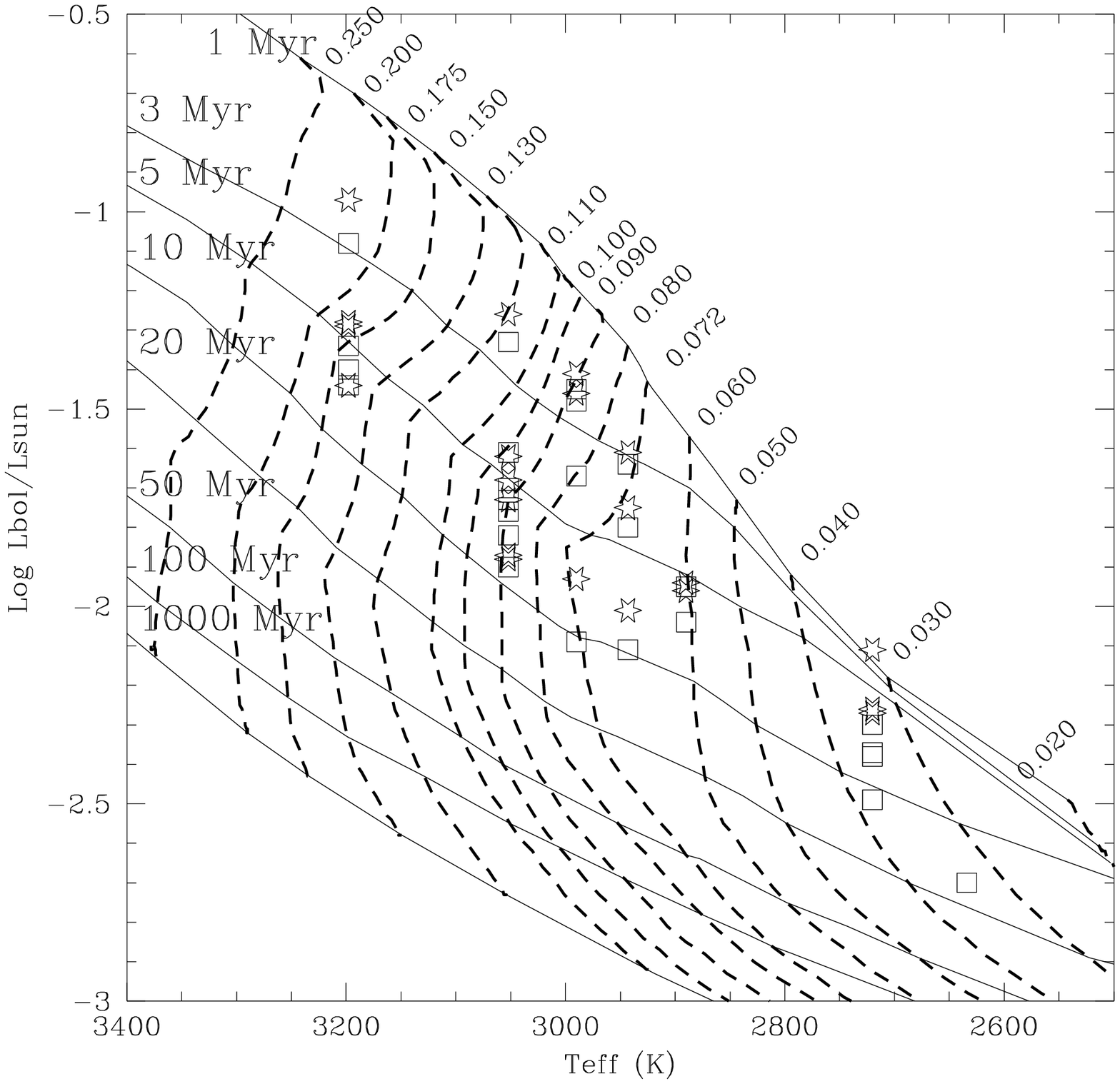}
 \caption{
{\bf a}
Dereddened color versus the absolute magnitude  for 
 Lambda Orionis candidate members --subsample with low
resolution spectroscopy.
The thin lines correspond to Baraffe et al. (1998) 1, 3, 5  and 10 
Myr isochrones. Solid circles correspond to probable members,
solid triangle to possible 
(''Mem+'' and ''Mem?'', respectively). 
{\bf b}  Hertzsprung-Russell diagram for Lambda Orionis
cluster probable and possible members
--subsample with low resolution spectroscopy. Open squares and stars 
correspond to bolometric luminosities  derived with $Ic$ and $Ks$
magnitudes (see text).
}
 \end{figure*}

\newpage

\setcounter{figure}{8}
    \begin{figure*}
    \centering
    \includegraphics[width=15.8cm]{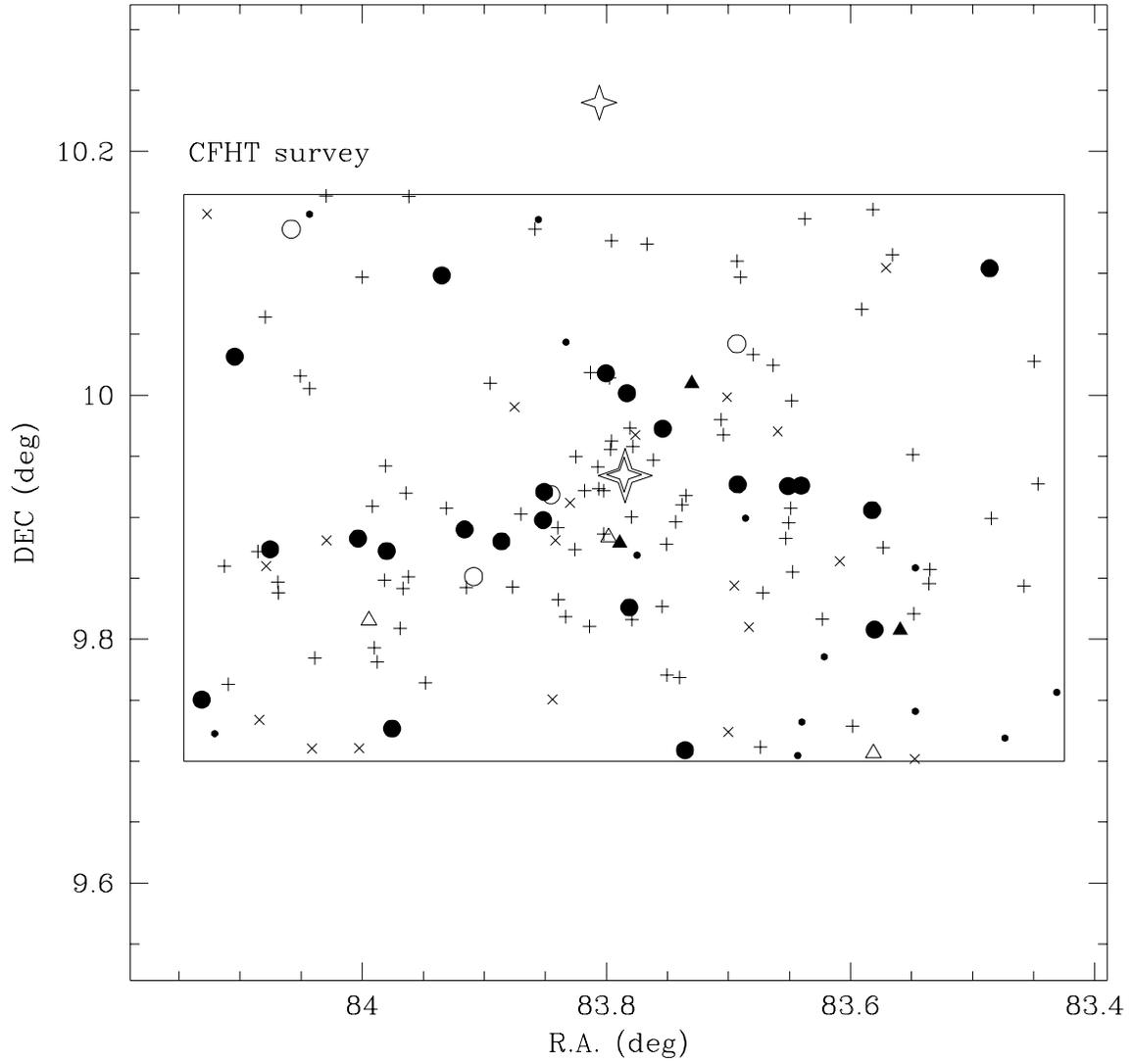}
 \caption{Spatial distribution of our candidate members.
 Symbols as in figure 5.}
 \end{figure*}

\newpage

\setcounter{figure}{9}
    \begin{figure*}
    \centering
    \includegraphics[width=9.8cm]{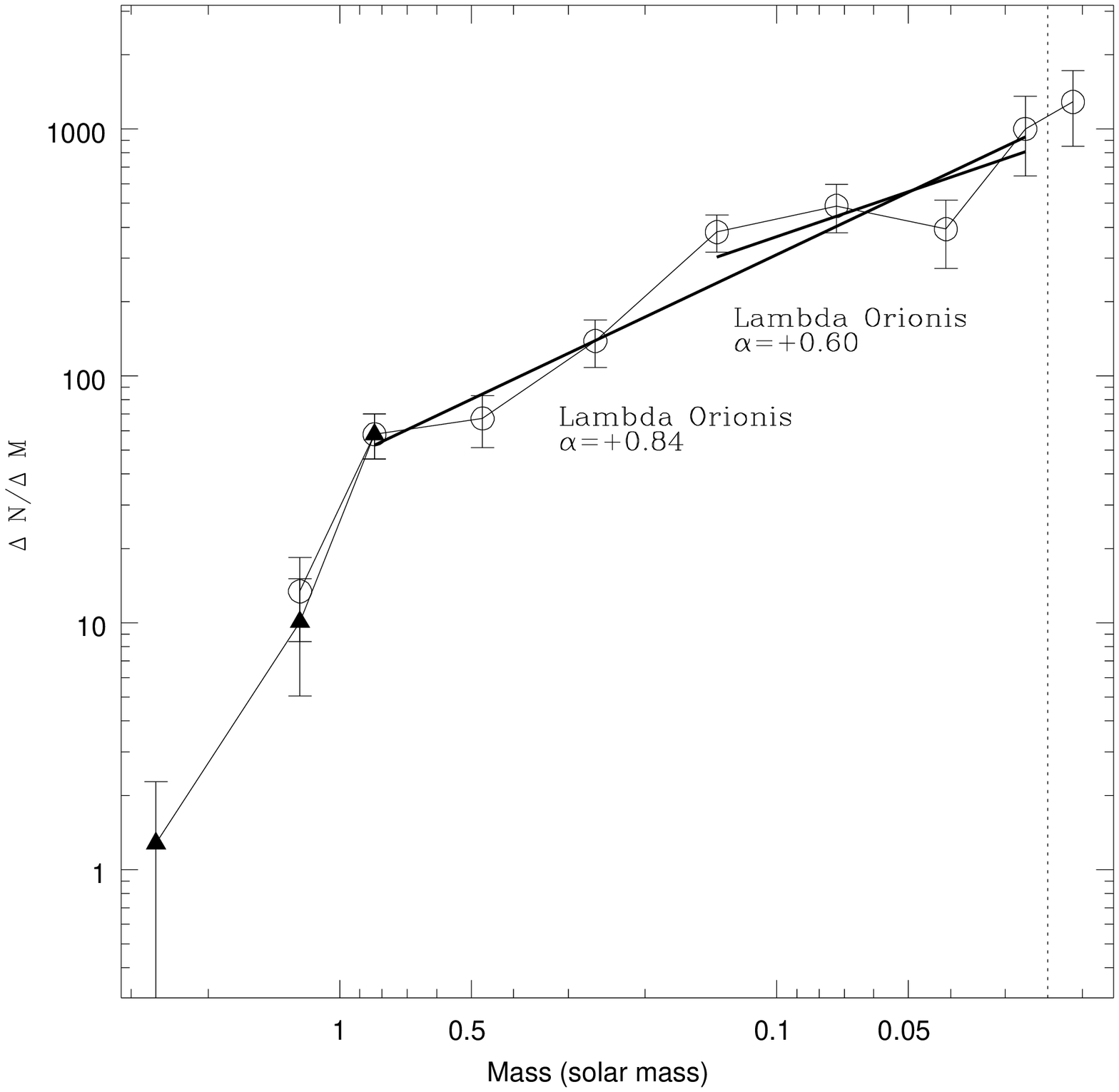}
    \includegraphics[width=9.8cm]{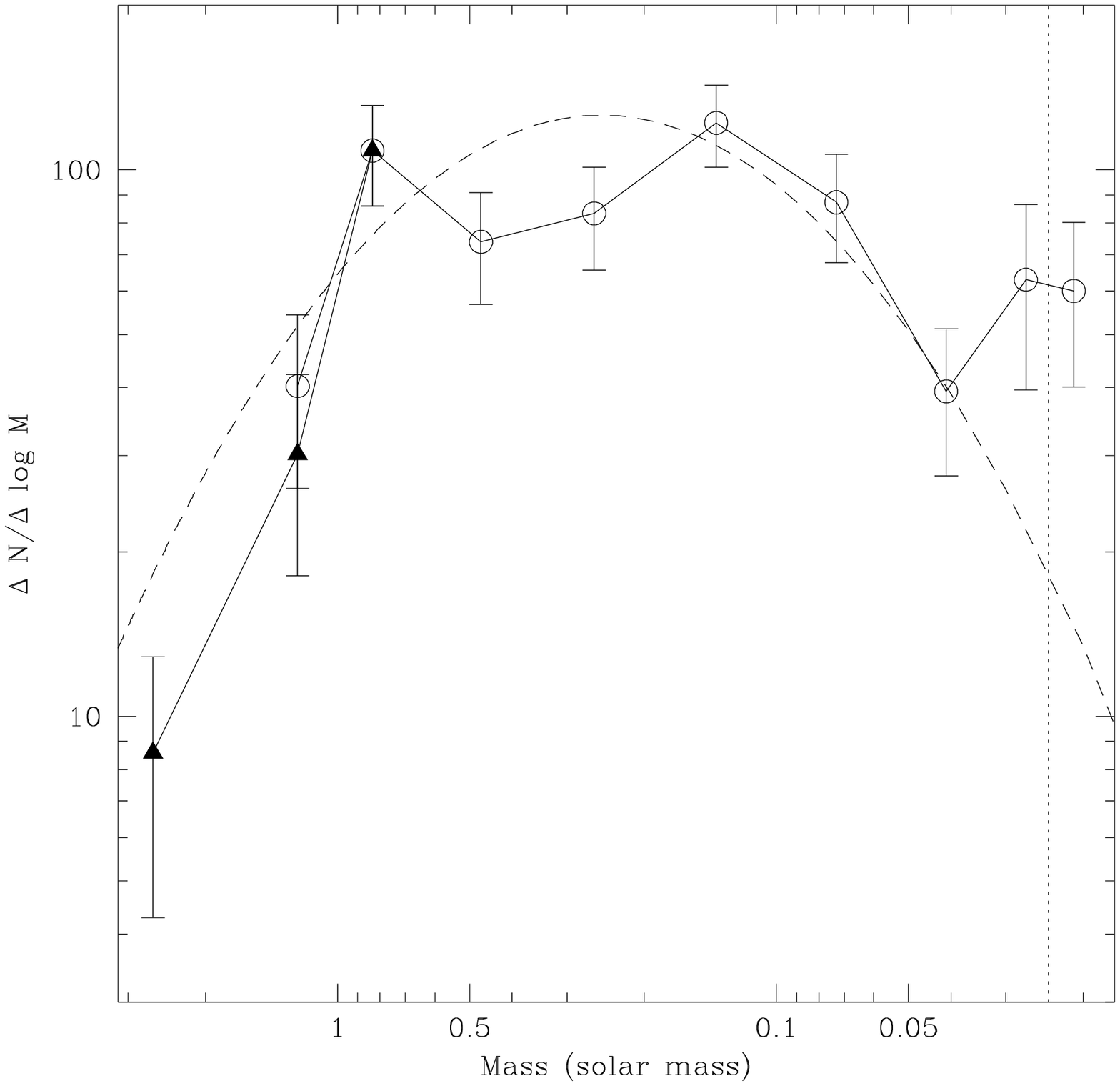}
 \caption{
Initial Mass Function for the Lambda Orionis cluster.
{\bf a} Solid triangles and open circles correspond to data from 
Dolan \& Mathieu (1999, 2001) and this work --same area, respectively.
The vertical segment corresponds  to the completeness limit.
{\bf b} Mass Function for Lambda Orionis -solid line- and the Pleiades
(dashed curve, Moraux et al. 2003) in lognormal form. }
 \end{figure*}


\newpage


\setcounter{figure}{10}
    \begin{figure*}
    \centering
    \includegraphics[width=15.8cm]{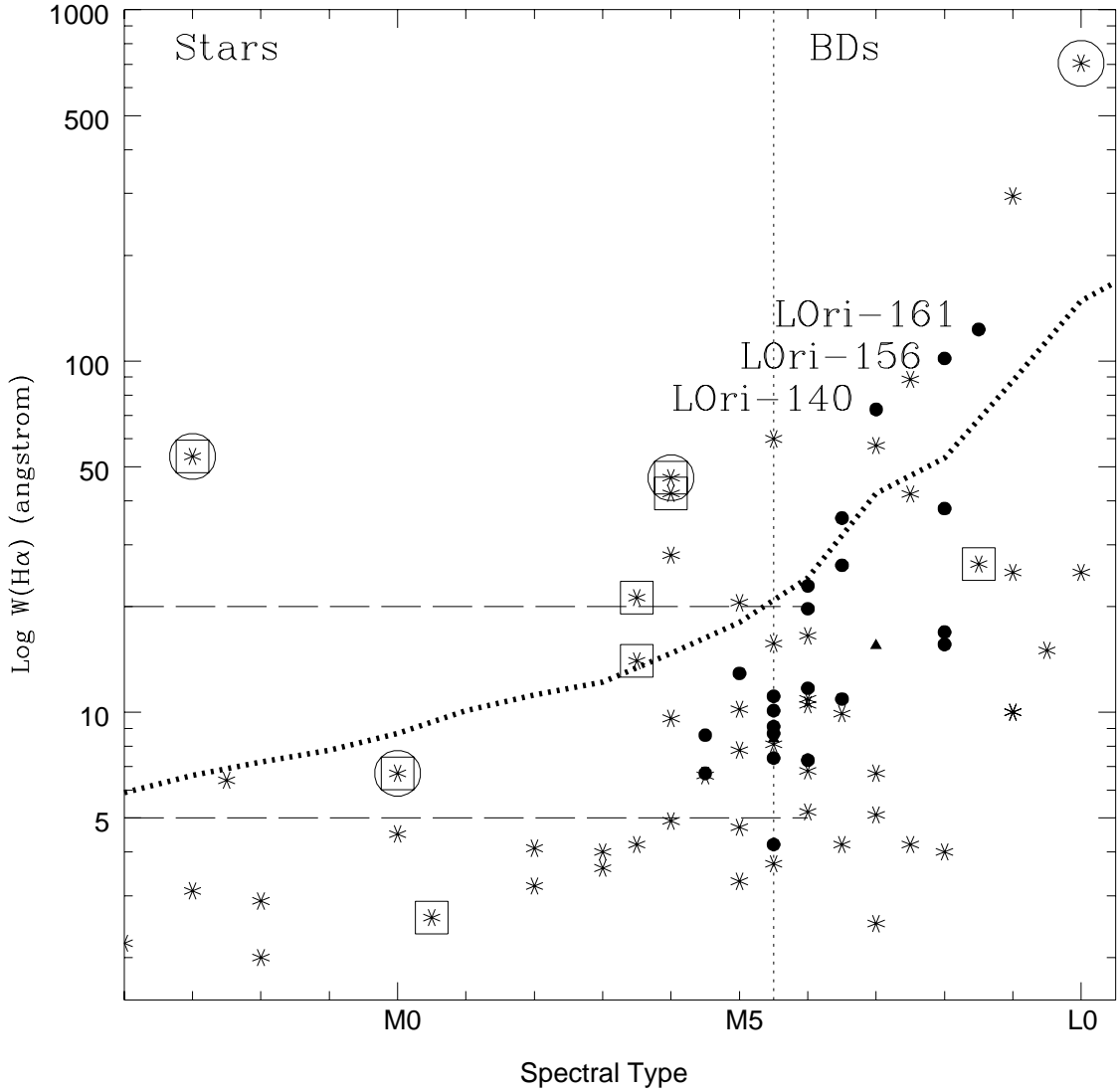}
 \caption{H($\alpha$) equivalent width versus the spectral type.
The values for Lambda Orionis candidate members 
(open circles for ''NM+'', solid triangles for ''Mem?'',
and solid circles for ''Mem+'') are compared with members of the ``twin''
cluster sigma Orionis (asterisks). Big circles and squares
indicate sigma Orionis members having infrared excesses and forbidden lines
in the spectrum, respectively. 
The dashed lines 
indicate two traditional criteria which separate Weak-line  from
Classical T~Tauri stars (5 and 20 \AA).
The light, vertical dotted line indicates the expected 
separation between stars and BDs at 5 Myr. 
 The thick dotted line corresponds to the
criteria differentiating accreting from non accreting objects,
 based on low 
resolution spectroscopy (Barrado y Navascu\'es \& Mart\'{\i}n 2003).
}
 \end{figure*}


\newpage

%
%



\end{document}